\documentclass[a4paper,12pt]{article}

\usepackage[utf8]{inputenc}
\usepackage[margin=3cm]{geometry} 
\usepackage[hyperfootnotes=false]{hyperref}
\usepackage{url}
\usepackage[nottoc]{tocbibind}
\usepackage{natbib}
\usepackage{enumitem}
\usepackage{graphicx}
\DeclareGraphicsExtensions{.pdf,.eps,.tiff,.png,.jpg}
\usepackage{setspace}
\usepackage{dcolumn}
\usepackage{wasysym}
\usepackage{booktabs}
\usepackage{colortbl}
\usepackage{amssymb}
\usepackage{fancyhdr}
\usepackage{subfiles}
\usepackage{lineno}
\graphicspath{{images/}{../images/}}

\title{\textbf{\thetitle}}

\date{}

\begin{document}

\pagenumbering{arabic}
\begin{center}
\Large  
\textbf{The cost of coordination can exceed the benefit of collaboration in performing complex tasks} 

\normalsize
\noindent \newline \\ Vincent J. Straub\textsuperscript{1}, Milena Tsvetkova\textsuperscript{1,2}, and Taha Yasseri\textsuperscript{1,3,4,5}* 
\end{center} 

\footnotesize 
\noindent  \\ \textsuperscript{1}Oxford Internet Institute, University of Oxford, Oxford OX1~3JS, UK \\
\textsuperscript{2}Department of Methodology, London School of Economics and Political Science, London WC2A~2AE, UK  \\
\textsuperscript{3}School of Sociology, University College Dublin, Dublin D04 V1W8, Ireland \\
\textsuperscript{4}Geary Institute for Public Policy, University College Dublin, Dublin D04 N9Y1, Ireland\\
\textsuperscript{5}Alan Turing Institute for Data Science and AI, 96 Euston Rd, London NW1 2DB, UK
\\ \newline
\noindent *To whom correspondence should be addressed: taha.yasseri@ucd.ie  
\normalsize 
\subsection*{Abstract}
Humans and other intelligent agents often rely on collective decision making based on an intuition that groups outperform individuals. However, at present, we lack a complete theoretical understanding of when groups perform better. Here we examine performance in collective decision-making in the context of a real-world citizen science task environment in which individuals with manipulated differences in task-relevant training collaborated. We find 1) dyads gradually improve in performance but do not experience a collective benefit compared to individuals in most situations; 2) the cost of coordination to efficiency and speed that results when switching to a dyadic context after training individually is consistently larger than the leverage of having a partner, even if they are expertly trained in that task; and 3) on the most complex tasks having an additional expert in the dyad who is adequately trained improves accuracy. These findings highlight that the extent of training received by an individual, the complexity of the task at hand, and the desired performance indicator are all critical factors that need to be accounted for when weighing up the benefits of collective decision-making. 

\footnotesize
\vspace{1cm}

{\bf Keywords:} collective intelligence $|$ competence $|$ team $|$ decision-making $|$ citizen science

\normalsize
\onehalfspacing

\section*{ \large Significance statement}
Collaboration is one of the fundamental processes in humans’ social lives. With the invention of digital communication technologies designed to further facilitate collaboration, identifying the keys to successful collaborations is more desirable than ever. Previous work in this area consists primarily of lab experiments and therefore suffers from a lack of generalizability. In this work, we utilize a real world citizen science platform to run experiments with subjects recruited from a diverse pool of non-expert participants. We show that when the task is complex, making decisions as an individual can be better than joint decision-making in dyads, particularly where coordination requires time and effort. Our work can inform the design of collaboration platforms and advance the science of teamwork.
 
\section*{Introduction}
\doublespacing 
Intelligent agents, be they natural or artificial, constantly have to make decisions to solve complex problems. Humans are no exception; decision-making is omnipresent in socioeconomic life, occurring in households, classrooms and, increasingly, online \citep{Tsvetkova2016}. The common belief is that group decisions are superior to decisions made by individuals. The proverb “two heads are better than one” captures the intuition that two (or more) people working together are indeed more likely to come to a better decision than they would if working alone \citep{bahrami2010,Koriat2012}. To test this notion empirically, we experimentally study collective image classification tasks using the real-world task environment of a citizen science project, where participants classified pictures of wildlife, taken as part of a conservation effort, as accurately and quickly as possible.

Our study focuses on dyadic interaction and was designed to reveal to what extent task complexity and variation in learned expertise influence the accuracy of classifications.  We ask three core questions: whether and when dyads perform better than an average individual, whether dyads with similar training perform better than dyads with different training, and how performance is mediated by the complexity of the task.

Surprisingly, we find little support for two heads being more accurate than one except for the most complex tasks, for which having an additional expert significantly improves performance upon that of non-experts. Our results show that pairs of individuals gradually improve in performance as they work together but tend not to experience a collective benefit compared to individuals working alone; rather, the cost of coordination to efficiency and speed is consistently larger than the leverage of having a partner, even if they are expertly trained. 
These findings highlight that the extent of training received by an individual, the complexity of the task at hand, and the desired performance outcome are all critical factors that need to be accounted for when weighing up the benefits of collective decision-making. 

Our findings stand in stark contrast to a vast literature of research on decision-making showing that groups usually make better decisions than individuals; most of that research considers groups with more than two members. Beginning with the discovery of judgment feats achieved by large numbers of people, classically in point estimation tasks like guessing the weight of an animal \citep{Galton1907}, the idea of the “wisdom of the crowd” \citep{Surowiecki2004} has become a prominent example. Building on other cases from stock markets, political elections, and quiz shows, evidence from other guessing tasks and problem-solving experiments \citep{Kerr2004,Grasso2012} shows that the aggregate of many people's estimates often tend to be closer to the true value than all of the separate individuals. 

Recent examples of research on the wisdom of the crowd include answering general-knowledge questions \citep{Navajas2018} and estimating political events \citep{Becker2019}. Yet, while some have found the effect holds for higher-dimensional tasks involving spatial reasoning and combinatorial problems such as the traveling salesman task \citep{yi2012}, in general, most are different to our study as they nevertheless continue to rely on numeric judgment tasks, such as dot estimation \citep{Almaatouq2020}. This difference in task context means most studies have focused simply on understanding the collective decision making through the pooling of personal information, often via simple averaging. 

Importantly, as studies have turned their attention to understanding what determines the performance advantage of collective decision-making, they have reached divergent conclusions on the importance of (i) diversity, in terms of individual team member attributes \citep{Hong2004,DeOliveira2018}; (ii) group size and structure \citep{Galesic2018,Navajas2018}; (iii) incentive schemes \citep{Mann2017}; (iv) the nature of the task, also referred to as the “task context” (used henceforth) or “task environment” \citep{Mata2012}; and, perhaps most contentiously, (v) the role of social influence, that is, whether interaction undermines \citep{Lorenz2011,Muchnik2013} or improves \citep{Becker2019,Farrell2011} collective performance. 

With respect to pairs of individuals (dyads)—the most elementary social unit \citep{Kozlowski2013} and focus of this paper—related recent experimental studies have tended to examine the role played by (1) individual confidence \citep{Koriat2012,Bang2014} and (2) argument quality \citep{trouche2014} in determining the success of collective decision-making. When taking others’ opinions into account on various numerical estimation tasks, studies have shown that people are known to rely on a ‘confidence heuristic’ \citep{Yaniv2007} such that more confident opinions tend to weigh more. For instance, \cite{Koriat2015} has shown that in the case of perceptual or even general knowledge questions, the outcome of group discussion can be emulated by aggregating the individual judgments of the group members weighed by their confidence. Meanwhile, studies of intellective tasks, problems or decisions for which there exists a demonstrably correct answer within a verbal or mathematical conceptual system, have focused on the idea that arguments, more than confidence, explain the good performance of reasoning groups \citep{trouche2014}. \cite{blanchard2020} report that decision outcomes improved when people act in dyads compared with acting individuals to complete a general-knowledge test. 

Recent studies have reported on other challenges in dyadic collaboration. The dyad members can suffer from an egocentric preference for personal information and fail to reap collective benefits by not listening to their expert counterparts \citep{Yaniv2007,Tump2018}. Similarly, pairs may underperform due to people’s tendency to give equal weight to others’ opinions \citep{mahmoodi2015}. Finally, individuals experience a collective benefit only when dyad members have similar levels of expertise and training, as this ensures members accurately convey the strength of their belief and the dyad can reliably choose the best strategy \citep{Bang2017,Bang2014}. 

Based on the observation that such critical agent characteristics, particularly the relevance of prior personal knowledge (i.e. expertise), are not randomly distributed, our work builds on this line of work by directly operationalizing differences in the extent of task-relevant training between interacting individuals—in contrast to studies which have relied on proxies for fixed differences in decision experience and ability \citep{Sella2018}.

Crucially, most prior studies typically provide participants with immediate feedback, alongside treating collective benefit as a static event. However, this fails to take into account that actual human pairs often have to perform tasks under uncertainty, whilst simultaneously benefiting from continued individual and social learning. The present study narrows the gap between experimental control and realistic settings by examining collective decision-making in the context of an established citizen science task—something that has not hitherto been tried for large-scale image classification platforms. 

\subsection*{Study Design}

\subsection*{The Wildcam Gorongosa task}
In the experiment, participants had a set amount of time to classify pictures taken by motion-detecting camera traps in Gorongosa National Park in Mozambique as part of a wildlife conservation effort (the trail cameras are designed to automatically take a photo when an animal moves in front of them). Specifically, we instructed participants, physically present in the lab and seated in front of a computer monitor, to use the Wildcam Gorongosa website, hosted by The Zooniverse\footnote{ \url{https://www.zooniverse.org/projects/zooniverse/wildcam-gorongosa}.}, the world’s largest online platform for citizen science \citep{Cox2015}. To increase the generality of the findings, the exact number and sequence of images are not controlled at the individual level, as is the case in many prior studies that have employed abstract, stylized tasks. Rather, to reflect the real-world nature of the tasks used in the experiment, we ensured participants classified images in a manner consistent with the experience of volunteering citizens visiting the Wildcam Gorongosa site. 

\begin{figure}[!t]
\centering
\includegraphics[width=\textwidth]{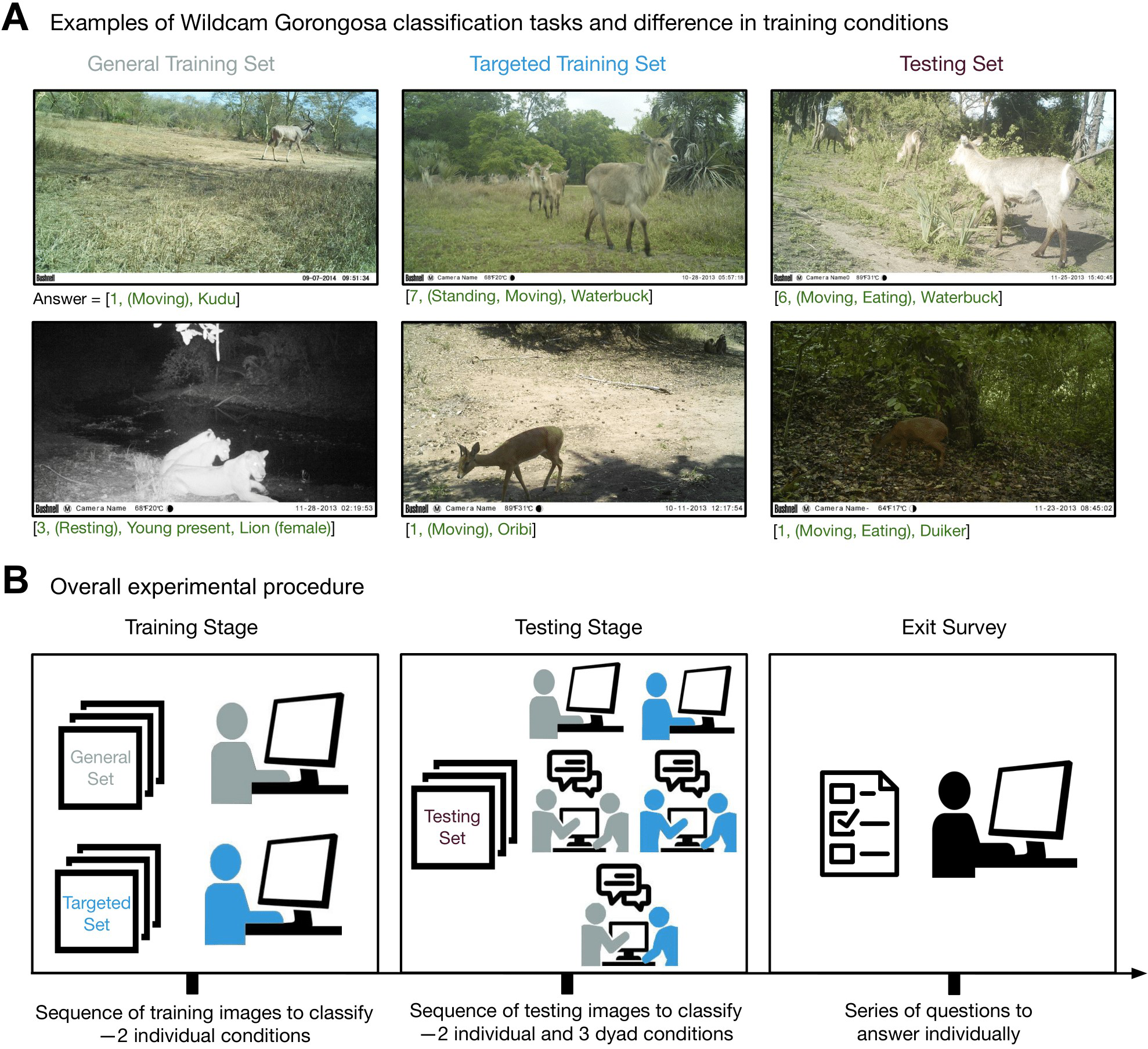} 
\caption{\small\textbf{Experimental design}. (\textbf{A}) Examples of the experimental task and illustration of the difference between the set of images seen in the general versus the targeted training condition. Instances of images classified in the testing stage are also provided. (\textbf{B}) Sequential schema of events in the Experiment. In the training stage, $T_1$ every participant classified images individually. In the testing stage $T_2$, participants were additionally assigned to either an individual or dyad condition. Wildcam Gorongosa imagery is reprinted under a CC BY 4.0 license with permission.}\label{fig:design}
\end{figure}

The Wildcam Gorongosa task consists of five distinct labeling subtasks which each involve  classifying an image with the correct label. Specifically, for each image, participants in the experiment had to perform the following classifications, listed in increasing complexity: (1) detecting the presence of the animal(s), (2) counting how many animals there are, (3) identifying the behaviors exhibited, specifically, identifying whether the animal(s) is (a) standing, (b) resting, (c) moving, (d) eating, or (e) interacting (multiple behaviors may be selected), (4) recognizing whether any young are present, and (5) identifying the species type. The 52 possible species options include a `Nothing here' button but no `I don’t know' option (see Fig.~\ref{fig:design}A for example images and Supplementary Material Figs.~S1-S4 for further examples and screenshots of the online platform and instructions). 

Prior work has demonstrated that object recognition is a task that requires context-dependent knowledge and various facets of our visual intelligence \citep{DiCarlo2012}; hence, citizen scientists primarily carry out tasks for which human-based analysis often still exceeds that of machine intelligence \citep{Trouille2019}. In the present case, the Wildcam Gorongosa task is thus considered suitable for the purposes of analysis for the following two reasons. First, it has high ecological validity: as part of an established crowdsourcing platform, a type of ``human-machine network" characteristic of our hyper-connected era \citep{Tsvetkova2017}, the site has engaged over 40,000 volunteers to date \citep{WildcamGorongosa}. Given the abundance of situations in everyday life where immediate outcomes are difficult, sometimes even impossible, to establish, the task can also be generealized to other contexts in that no feedback was provided to participants. Second, identifying various features in camera trap wildlife images is sufficiently difficult \citep{Norouzzadeh2018} that it offers the possibility of collaborative benefits, especially when dyad members have received similar training for identifying particular features.  

We operationalize task complexity experimentally by varying the number of information cues that subjects consider\citep{de2017understanding,groen2018scene}. In particular, as our definition means the complexity of each subtask is based upon the number of visual information cues that must be processed in order to succeed in the task, identifying the species is considered the most complex because it is expected to require processing the most inputs \citep{Swanson2016,parrish2018exposing}. 

Whilst this conceptualization diverges from classical organizational behavior definitions \citep{Wood1986, hackman1969toward}, which also consider the actor doing the task, the task context, and the behavioral pattern required to perform the task, it has parallels with the notion of ‘component complexity', part of  the original framework proposed by \cite{Wood1986}. We thus nevertheless retain the original label `task complexity` because our definition retains a connection to the original concept. Differences in learned expertise are meanwhile operationalized as differences in task-relevant training, which is taken to be the main way to facilitate the acquisition, enhance retention, and promote the transfer of relevant knowledge to contexts not encountered during training \citep{healy2014training}.

\subsection*{Experimental Procedure}
\begin{figure}[!t]
\centering
\includegraphics[width=0.7\textwidth]{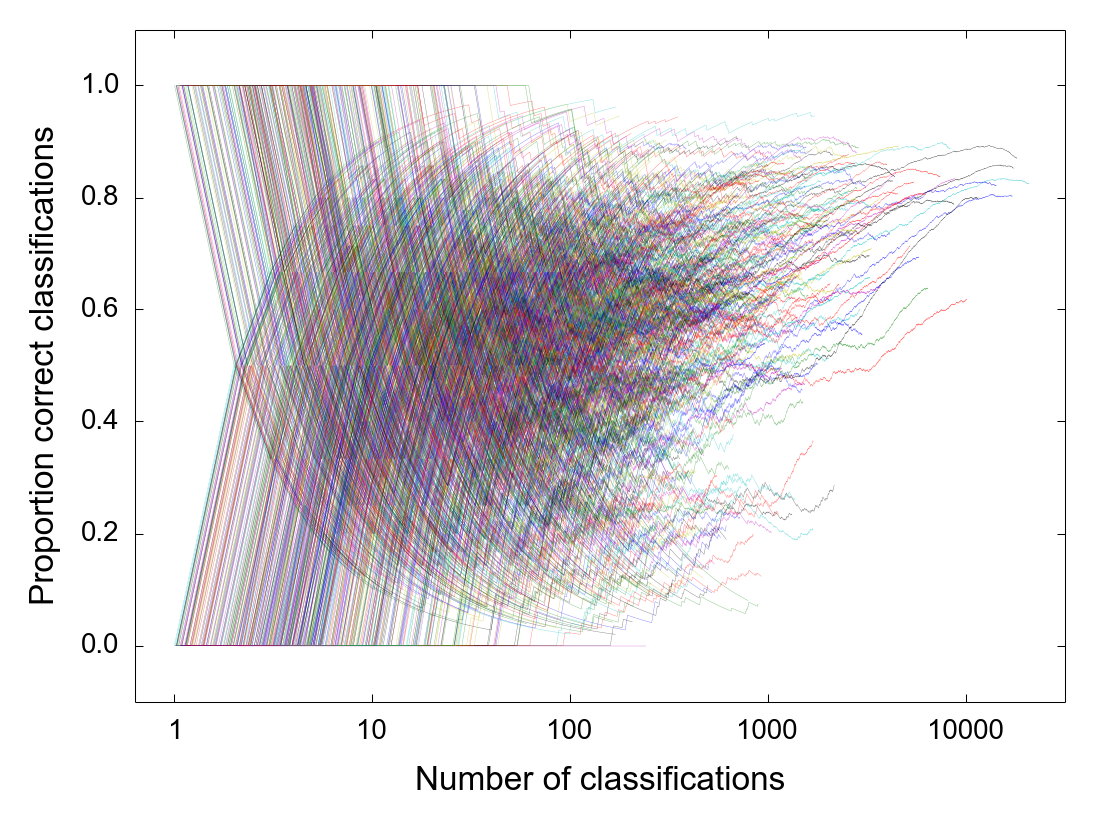} 
\caption{\small\textbf{Individual learning trajectories of citizen scientists}. Trajectories are defined as the change in the proportion of correct classifications averaged across each of the five Wildcam Gorongosa tasks relative to the number of classifications made. Each learning trajectory is for an individual volunteer citizen scientist who accessed the Wildcam Gorongosa site prior to the time of the experiment. Note the logarithmic scale on the x-axis, hence for a small number of classifications, the curves take exotic shapes.}\label{fig:trajectories}
\end{figure}

The experiment was divided into a training and a testing stage ($T_1=30 ~\rm{min}$ and $T_2=45 ~\rm{min}$). A training stage was required in order to ensure select participants could receive task-relevant training and build up greater expert knowledge in performing the task at hand; this in turn ensured mixed-skill dyads could be formed for the testing stage where one member (the `expert') has higher learned expertise. The length of the training stage was determined by analyzing pre-existing citizen scientist-generated data for the Wildcam Gorongosa task. In particular, we analyzed the `learning trajectories' of users, and changes in performance over time, to determine the average saturation time point, defined as the point after which the majority of users no longer improve. As depicted in Fig.~\ref{fig:trajectories}, most users who classified the first image correctly (incorrectly) made an incorrect (correct) classification within 10 classifications; only a small subset maintained their winning (losing) streak. After determining the average number of classifications an individual made before no longer experiencing an increase in the proportion of classifications classified correctly and multiplying this count by the average time it took to make a single classification, this was found to be 30 minutes. 

For the $T_1$ period, participants were randomly assigned to two different training conditions, `General' and `Targeted', and asked to individually classify a sequence of approximately 50 images, where the content of the images was varied between conditions. As $T_1$ was designed to provide selective training to participants in the Targeted treatment condition, the set of images seen by these participants (Targeted set) consisted of pictures sharing specific features, sampled from a predetermined subset of all the images available to classify on Wildcam Gorongosa. Specifically, the Targeted set was restricted to pictures containing antelope species: bushbuck, duiker, impala, kudu, nyala, oribi, reedbuck, and waterbuck. These animals were chosen as they look visually similar, share a number of morphological features, and exhibit similar behaviors, thus making them relatively harder to distinguish \citep{Norouzzadeh2018}. For the General condition, pictures shared less specific features and were instead sampled from a much broader predetermined subset containing a more diverse set of animals other than the ones mentioned above which are easier to distinguish, including baboons, warthogs, and lions, among many others (see Supplementary Text ~S2.3.1 for details). The analysis presented below confirms the effectiveness of this treatment in producing significant differences between different groups.

$T_2$ was designed to assess the effect of differences in task-relevant training as well as the effect of working alone versus collectively. For this reason, participants were further randomly assigned to either a `Solo' or `Dyad' condition, with participants in the dyad condition having to make a joint decision by reaching a consensus (as opposed to voting or relying on an averaging procedure); the dyad members worked together and one of them was randomly assigned to input decisions on behalf of the pair. When taking into account the level of training, the experimental procedure resulted in 2 individual testing conditions, `General Solo' and `Targeted Solo', and 3 dyad testing conditions, `General Dyad', `Targeted Dyad', and `Mixed Dyad'. The set of images seen by all participants, regardless of testing condition, thus consisted of pictures sampled from the Targeted set (see Supplementary Text ~S2.3.2 for details). Fig.~\ref{fig:design} illustrates the overall experimental design. 

Each decision process that we measure can be broken down into two phases: (1) individual or (in the case of dyads) joint deliberation, deciding what decision to make, and (2) execution, confirming and inputting the decision using the computer mouse. Because we consider collective decision-making to be primarily concerned with the process outcomes of interaction, we are interested in the former phase: how effectively dyads deliberate (i.e., share social information). In the execution phase, dyads may be expected to be, at least initially, slower in inputting their decision as they have to coordinate their classification. However, we consider this to have a negligible impact on their overall performance and expect it to diminish over time as they learn to coordinate. Instead, we expect their performance to be determined primarily by how effectively they interact during deliberation, so we do not try to arbitrarily discount the effect of the execution phase on performance via any type of weighting. 

To measure performance, three metrics were in turn computed for every individual or dyad $i$: \textit{pace}, defined as the number of images classified by $i$ per minute (referred to as \textit{volume} when considering the absolute number); \textit{accuracy}, defined as the number of correct guesses made by $i$ as a proportion of volume; and \textit{efficiency}, defined as the number of correct guesses made by $i$  per minute. These metrics are chosen as they have been widely used as measures of performance in decision science \citep{Maloney2002}. Except for pace, which is an aggregate measure---and thus the same across tasks---accuracy and efficiency were computed separately for the five tasks (see Supplementary Text ~S3.2 for details). 

\section*{Results}

\begin{figure}[t!]
\centering
\includegraphics[width=0.75\textwidth]{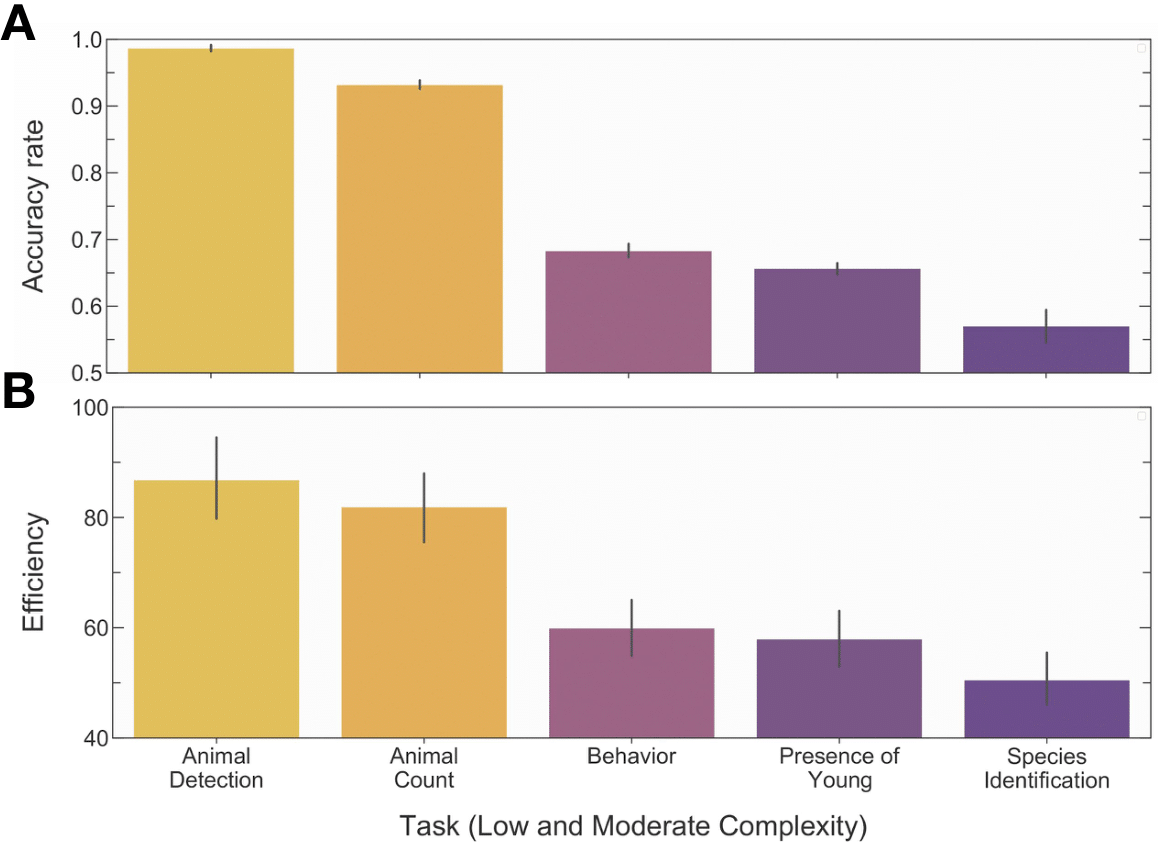} 
\caption{\textbf{Task complexity mediates performance.} Data is combined across all solo and dyad grouping conditions for the testing stage. The more `complex' the task, the greater the reduction in average accuracy (\textbf{A}) and average efficiency (\textbf{B}). The difference in experienced difficulty between the task with the lowest and highest complexity is very large: the average accuracy score dropped by nearly 50\%. Error bars indicate the 95\% confidence intervals.}\label{fig:complexity}
\end{figure}

Fig~\ref{fig:complexity} shows the overall accuracy and efficiency of participants for different tasks. Tasks vary in their complexity, and therefore in the efficiency and accuracy of classifications. Based on these results we categorize detection of an animal and animal count as {\it low complexity} tasks and identifying the behavior, recognizing young animals, and identifying the species as {\it moderate complexity} tasks.

\subsection*{The effect of training on classifications}

Focusing first on the effect of training on performance across the different tasks of varying complexity, we assess the difference in performance changes over the course of the entire experiment where change is relative to an individual's initial performance (see Materials and Methods and Details of Analysis in the Supplementary Information for details). $T_1$ and $T_2$ were separately split into three equally spaced and non-overlapping time intervals. The average change in performance was then estimated separately for each interval by computing the difference in performance compared to the performance of the respective individual(s) in the first interval of the training stage, which can be thought of as the natural baseline prior to training. The rationale is that benchmarking against this initial performance allows for a more precise estimation of whether individual training, interaction (in the case of dyads) or both provide performance gains \citep{almaatouq2021task}. 

\begin{figure}[t!]
\centering
\includegraphics[width=1\textwidth]{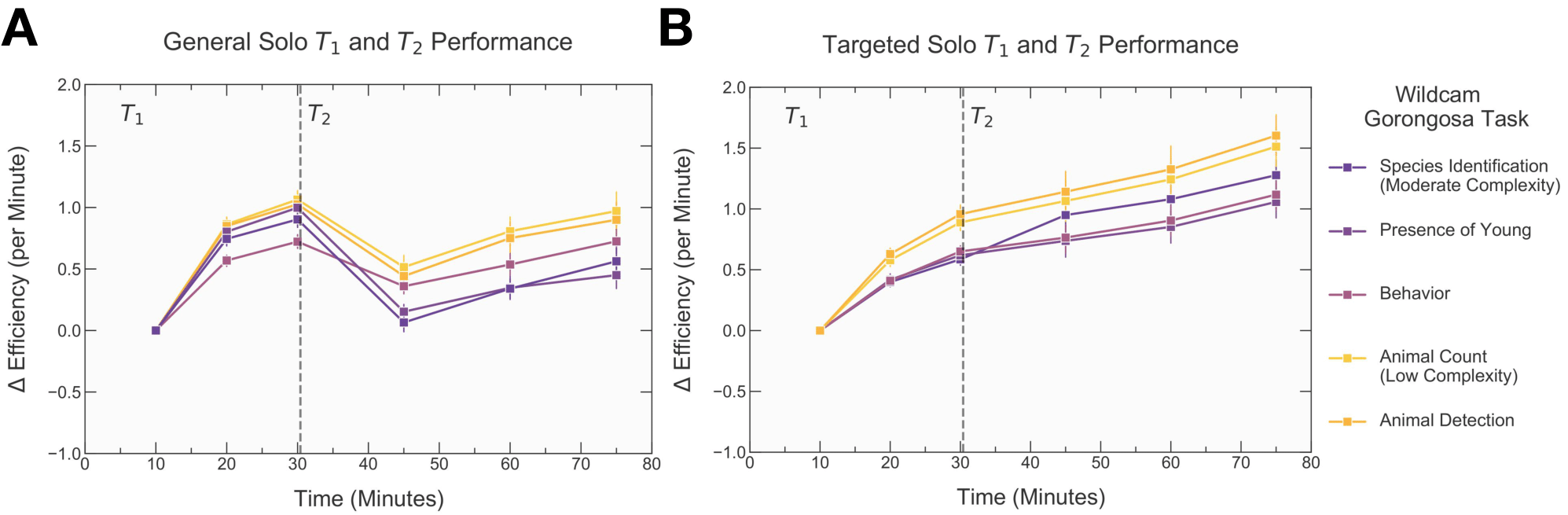} 
\caption{\small\textbf{Individual training effectiveness.} Performance changes in terms of the average change in efficiency across each of the five Wildcam Gorongosa tasks during the training ($T_1$) and testing stage ($T_2$) for General Solo (\textbf{A}), individuals who received general training during $T_1$, and Targeted Solo (\textbf{B}), individuals who received selective training. The dashed vertical line falling in $T_2$ separates both stages, which each have 3 data points. Error bars indicate one standard error of the mean.}\label{fig:results}
\end{figure}

Fig.~\ref{fig:results} depicts the results, showing the average change in efficiency across each of the five Wildcam Gorongosa tasks during the training ($T_1$) and testing stage ($T_2$) for participants in the General Solo (Fig.~~\ref{fig:results}A) and Targeted Solo (Fig.~~\ref{fig:results}B) groups. Relative differences in learning rates and performance changes indicate that individual performance is mediated both by the level of training received and the complexity of the task \citep{Saffell2003,Mao2016a,Mata2012}. 

Individuals in the General Solo group continuously improve across each task during $T_1$, becoming most efficient during the last third of training, as the benefits of general training gradually subside, before experiencing a sharp decline during the first phase of testing. Although they are able to recover by the end of $T_2$ with respect to animal detection, count, and behavior---tasks of lower complexity---the consistency of the decline across tasks indicates that individuals with general training are initially less efficient when confronted with new task stimuli. In contrast, individuals in the Targeted Solo group continuously improve upon their $T_1$ performance throughout the course of $T_2$, experiencing greater relative gains in efficiency compared to General Solo for each time interval (Fig.~\ref{fig:results}B), suggesting that targeted training provides consistent accumulative performance benefits.

\subsection*{Individual versus collective classifications}

We now turn to our primary research questions: whether dyads outperform individuals and how that the performance of dyads depends on task complexity and individual ability. We analyze differences in performance change during the final phase of the testing stage, when the benefits of interaction for dyads can best be separated from additional effects, such as initial improvements in coordination. As a secondary analysis, we also examine differences over the course of the entire testing stage; this allows us to see how these changes evolve, which in turn allows us to detect when and why coordination problems may arise, for instance, as a result of an initial conceptual shift. 

The performance of Mixed Dyads is computed in two ways: using a `General benchmark' (the performance set by the member with general training) when comparing Mixed Dyads to General Solo and General Dyads, and using a `Targeted benchmark' (the performance set by the member with targeted training) when comparing Mixed Dyads to Targeted Solo and Targeted Dyads (see Material and Methods). 

\begin{figure}[t!]
\centering
\includegraphics[width=0.99\textwidth]{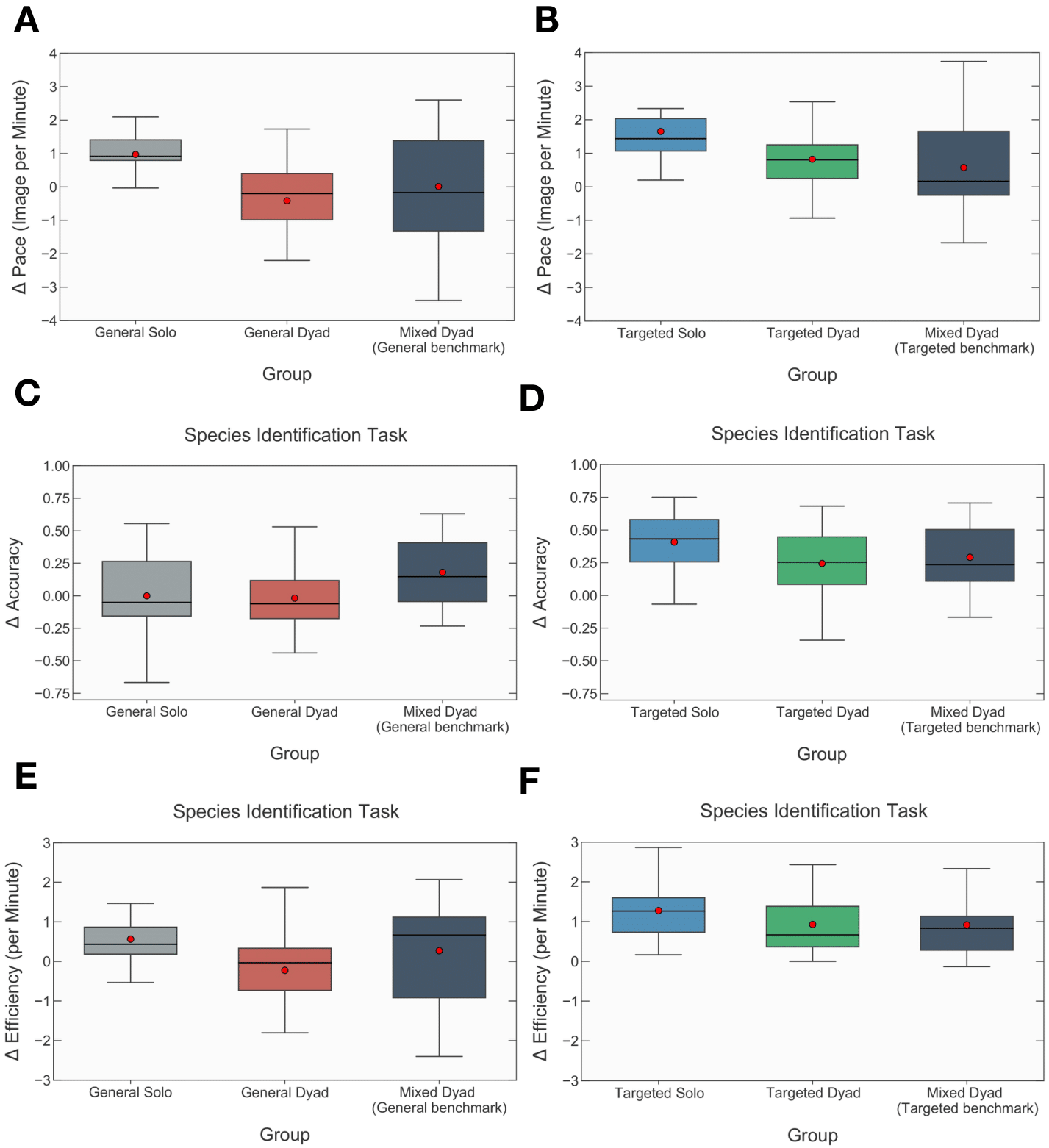} 
\caption{\small\textbf{Individual and collective performance}. Performance differences between individuals and dyads for the last third of $T_2$ in terms of the distribution and average change in pace (\textbf{A-B}), alongside accuracy (\textbf{C-D}) and efficiency (\textbf{E-F}) for the species identification task, considered the most complex. General Solo individuals are compared to General Dyads and Mixed Dyads (left panels) and Targeted Solo individuals are compared to Targeted Dyads and Mixed Dyads (right panels). Error bars indicate one standard error of the mean. Whiskers are 1.5 times the interquartile range. The solid black line inside each box indicates the median and the red circles do the mean.}\label{fig:results_2}
\end{figure}

Fig.~\ref{fig:results_2}A-B depicts the results for pace during the last time interval of $T_2$ for the species identification task (most complex task), indicating that participants in an individual condition were on average faster in classifying images belonging to the test set--regardless of whether or not they received targeted training. Moreover, only Targeted Dyads and, to a lesser extent, Mixed Dyads still improved in terms of pace during the final phase; however, both General Dyads and Mixed Dyads benchmarked against individuals receiving general training saw no or close to no further improvements on average. Statistical tests of the differences provide further evidence that participants in the solo conditions outperform participants in the respective dyad conditions ($p<$ 0.05 for pairwise comparisons; see Supplementary Material, Sec.~S3.3). 

Moving on to task-specific performance differences in accuracy and efficiency for the species identification task, considered the most complex task, General Dyads experience no interaction benefit. Yet, pairing an individual with general training and an individual with targeted training strongly counteracts this result: Mixed Dyads improve in accuracy by 37\% more than General Solo ($p$=0.036, Cohen's $d_s$=0.616, 95\% CI for Cohen's $d_s$: [0.034, 1.199]) and 55\% more than General Dyads ($p$=0.018, Cohen's $d_s$=0.895, 95\% CI for Cohen's $d$: [0.298, 1.492])---indicating that participants with general training can improve upon their expected individual performance by collaborating with a selectively trained partner who can supply task-specific expertise. 

Targeted Solo individuals meanwhile experienced a significantly greater average improvement in accuracy during the last interval of $T_2$ (41\%; Fig.~\ref{fig:results_2}D) compared to Targeted Dyads (24\%; $p$=0.026), and to Mixed Dyads (29\%; $p$=0.134), although to a less evident extent. As a result, they appear to consistently achieve the greatest efficiency (Fig.~\ref{fig:results_2}F), albeit only slightly. In contrast, whilst General Solo achieved greater efficiency compared to General Dyads as a result of increased speed (Fig.~\ref{fig:results_2}A), achieving 0.78 more classifications per minute ($p$=0.002, Cohen's $d_s$=0.934, 95\% CI for Cohen's $d_s$: [0.322, 1.546]), the slight difference between General Solo and Mixed Dyads (Fig.~\ref{fig:results_2}E; $p$=0.557) implies that individuals without selective training experience a speed advantage when working alone but derive an accuracy benefit when collaborating with a selectively trained teammate. 

Surprisingly, across all other tasks, there was no clear indication of substantive differences between individuals and dyads in terms of accuracy during the final phase of testing (see Supplementary Material). 

\subsection*{Differences in classifications over time}

\begin{figure}[t!]
\centering
\includegraphics[width=\textwidth]{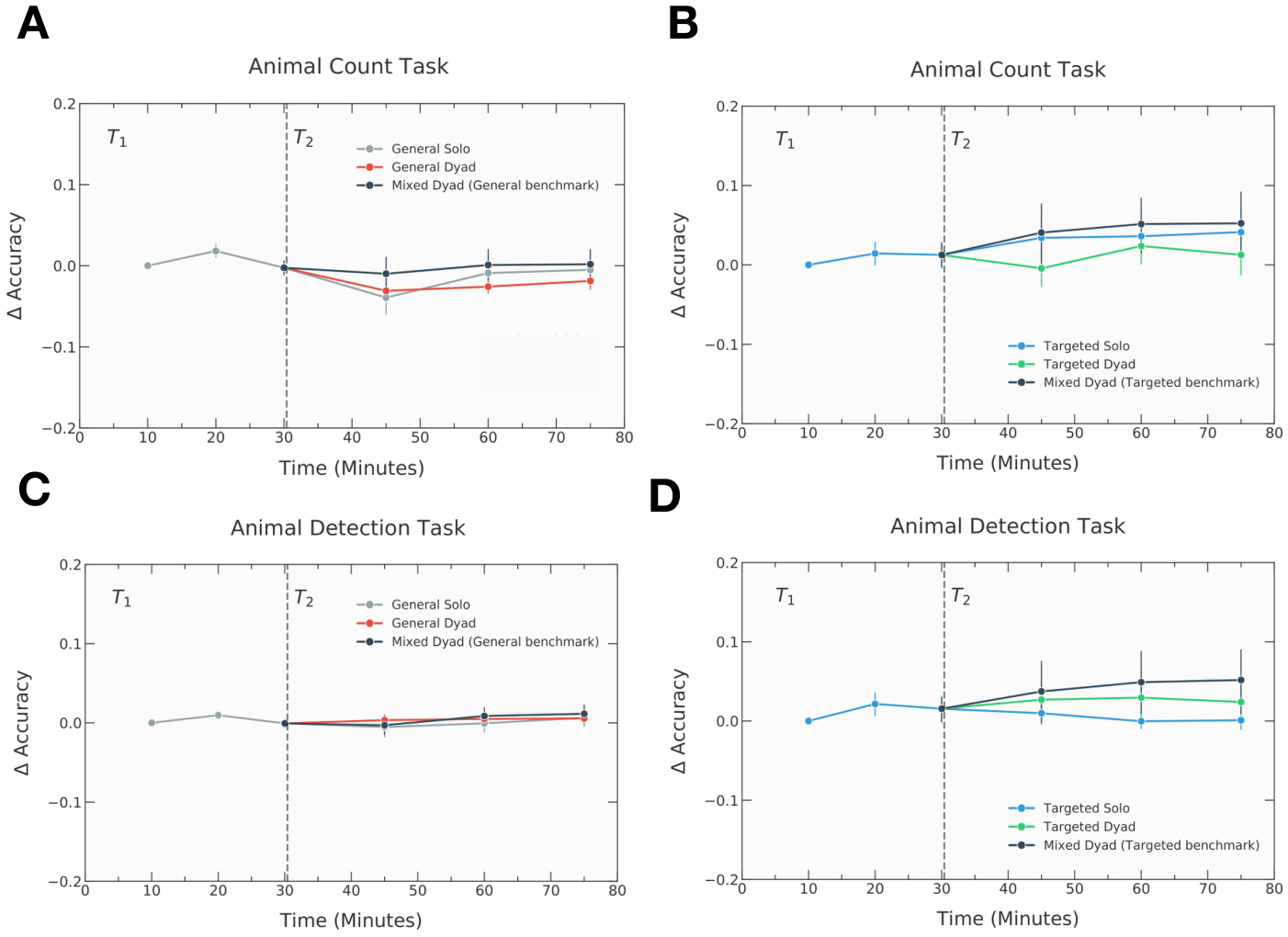} 
\caption{\textbf{Change in Accuracy over time over time}. Performance differences between individuals and dyads over the course of $T_2$ in terms of the average change in accuracy for low complexity tasks. Error bars indicate one standard error of the mean.}\label{fig:results3a}
\end{figure}

\begin{figure}[t!]
\centering
\includegraphics[width=\textwidth]{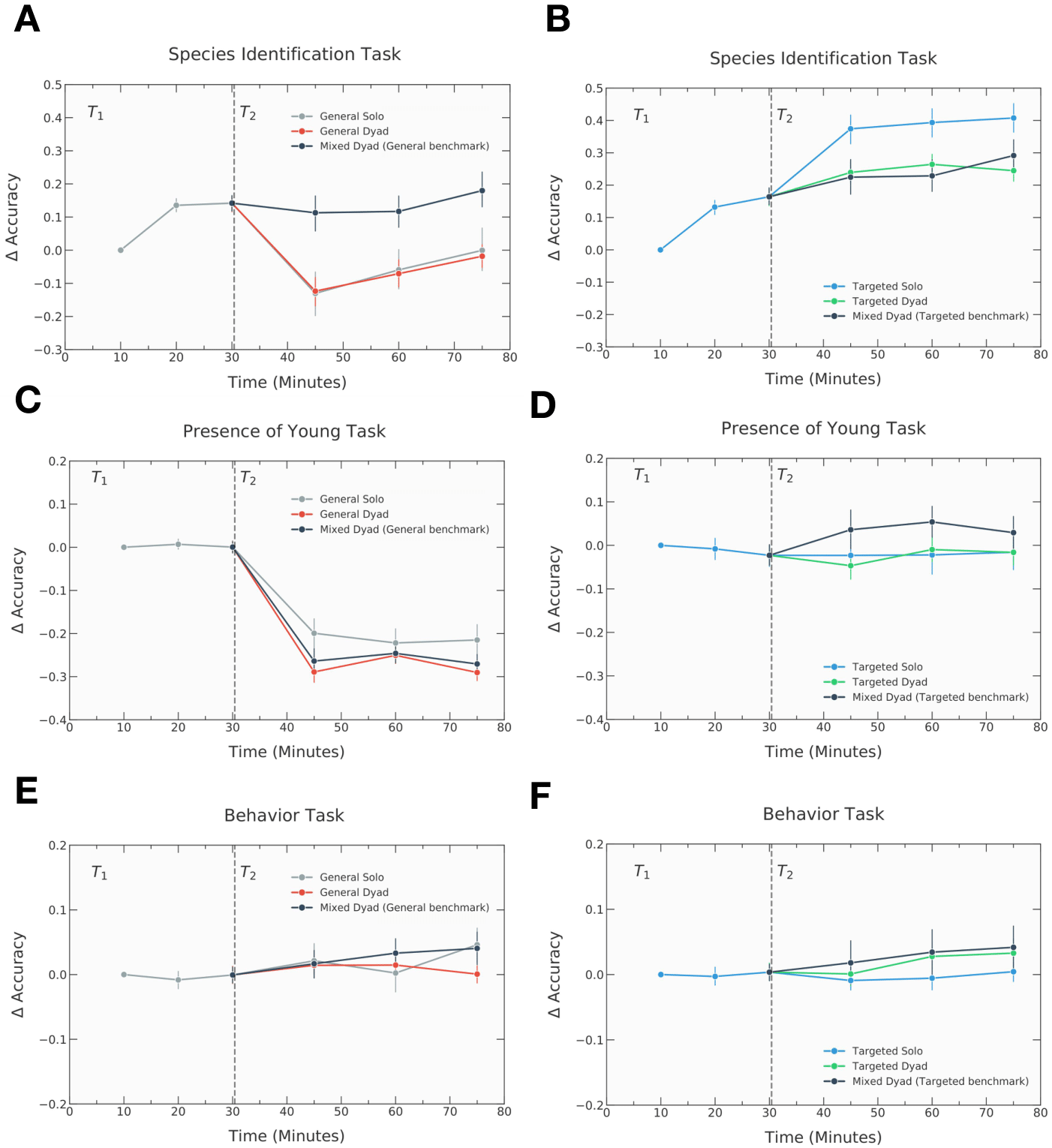} 
\caption{\textbf{Change in Accuracy over time over time}. Performance differences between individuals and dyads over the course of $T_2$ in terms of the average change in accuracy for medium complexity tasks. Error bars indicate one standard error of the mean.}\label{fig:results3b}
\end{figure}

Fig.~\ref{fig:results3a}~and~\ref{fig:results3b} show how accuracy changes over time for different tasks (grouped in low and medium complexity tasks respectively).

When comparing the change in accuracy over the course of the entire testing stage, there is no clear indication of significant or consistent differences between individual and dyads in terms of accuracy for less complex tasks; Fig.~\ref{fig:results3a} illustrates that this holds across the entire testing stage. When comparing changes in accuracy between General Solo and the respective dyad conditions for the presence of young task, considered to be of medium complexity (Fig.~\ref{fig:results3b}, a similar pattern can nevertheless be observed to changes in accuracy across the species identification task; notably, individuals and dyads both fail to experience any accuracy improvements. Moreover, when comparing Targeted Solo and the respective dyads conditions, Mixed Dyads appear to experience the greatest improvements in accuracy during the first two intervals of $T_2$. However, the slight variations suggest that tasks of low complexity are most probably too simplistic for practically relevant performance differences to emerge. 

\begin{figure}[t!]
\centering
\includegraphics[width=\textwidth]{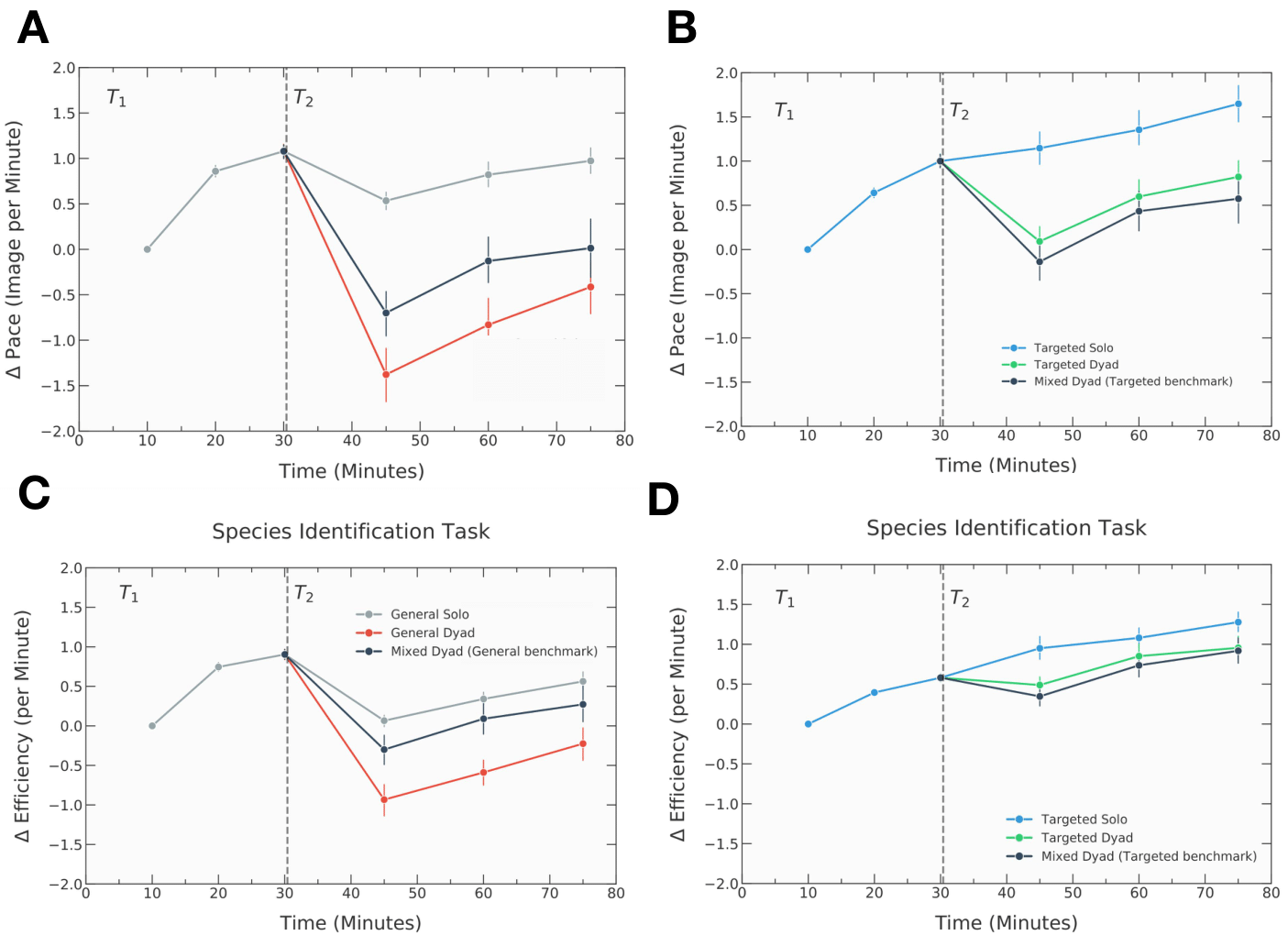} 
\caption{\textbf{Change in pace over time}. Differences between individuals and dyads over the course of $T_2$ in terms of the pace of task completion. Error bars indicate one standard error of the mean.}\label{fig:results3old}
\end{figure}

A decline is to be expected given the coordination issue, the extent of the decrease in performance experienced by dyads depends on the distribution of expertise among dyad members and has a different effect depending on this distribution. As seen in Fig.~\ref{fig:results3old}, whilst General Dyads and Mixed Dyads unsurprisingly undergo a decline in accuracy, pace and consequently efficiency, Targeted Dyads and Mixed Dyads only experience a fall in pace  but not accuracy, hence the slight dip in efficiency during the first time interval of $T_2$ (Fig.~\ref{fig:results3old}D) is primarily the result of classifying images more slowly. More specifically, individuals and dyads with general training experienced a reduction in accuracy during the first interval of $T_2$ 12\% and 13\%, respectively. 

Although both groups on average still gradually improved over time, they do not manage to recover to the performance benchmark set by General Solo during the first phase of training. As a consequence, the finding that General Dyads experience no interaction benefit at all remains consistent beyond the initial phase of testing. The observation that pairing an individual with general training and an individual with targeted training causes this decline is meanwhile consistently strong but becomes most pronounced in the final time interval. This not only implies that participants with general training can improve upon their expected individual performance by collaborating with a selectively trained partner but further suggests that the greatest interaction benefits among dyads with mixed expertise occur after overcoming initial coordination issues.

Regarding efficiency, Targeted Solo individuals appear to consistently achieve the greatest performance gains, as compared to Targeted Dyads and Mixed Dyads, albeit only slightly. In contrast, whilst General Solo achieved greater efficiency compared to General Dyads as a result of increased speed reflected in pace ($p$=0.018), the slight difference between General Solo and Mixed Dyads  ($p$=0.557) implies that individuals without selective training experience a speed advantage when working alone but derive an accuracy benefit when collaborating with a selectively trained teammate (see also Supplementary Material Fig.~S7, which shows that using different time intervals does not significantly change any of the reported findings). 

\section*{Discussion}

Our findings, in contrast to empirical studies of larger groups, demonstrate that the benefits of dyads are contingent. On the one hand, the stepwise performance improvements experienced by dyads in terms of pace and accuracy for the species identification task are consistent with prior studies showing that collective performance increases gradually for complex tasks in the absence of feedback \citep{Bahrami2012a,Mao2016a}. On the other, these findings imply that switching to a dyadic context for a task after training in an individual context in most cases creates a coordination problem that hurts performance. 

Although the observation that dyads do not consistently experience a collective benefit across tasks goes against prior results that indicate an interaction benefit in the context of low-level visual enumeration and numeric estimation tasks \citep{Koriat2015,Wahn2018}, this finding is nevertheless in line with literature that has found no pair advantage when the task context consists of general knowledge-based tests involving discrete choice decisions \citep{Schuldt2017,blanchard2020}. This underscores the extent to which performance greatly depends on overcoming any initial coordination issues and the task context. Whilst some previous studies have suggested that individuals outperform dyads due to faster skill acquisition \citep{Crook2010}, the greater accuracy gains achieved by Mixed Dyads when compared to General Dyads and General Solo for the species identification task shows that, at least for complex tasks, this may further depend on the level of task-relevant training among dyad members. 
Additionally, Our results highlight that task-relevant training, particularly targeted training, provides significant performance improvements, regardless of working individually or in a pair. 

Our work stresses that existing experiments and theories of collective decision-making, such as the ``confidence matching" model stating that only pairs similar in skill or social sensitivity experience a collective benefit \citep{Bang2017}, do not adequately stipulate if, and under what conditions, dyads will outperform individuals. 

The immediate contribution of this study is to demonstrate that in the context of a real-world task context, namely, Zooniverse's Wildcam Gorongosa citizen science project, the heads of two experts, understood here as individuals with targeted training, are not always better than one. Rather, one expert is more \textit{efficient} than two experts or a mixed ability dyad, indicating that the cost of coordination to efficiency is consistently larger than the leverage of having a partner---even if that partner is also specially trained for the task at hand.

Moreover, we also show that a non-expert, an individual with basic training, works faster than a dyad, even if one of the dyad members is an expert, thereby giving further support to the theory that dyads face a coordination problem which costs time and suggesting that individuals exert less effort when working in a pair. However, when it comes to \textit{accuracy} in the most complex task, having one expert in a dyad significantly improves performance compared to that of individual non-experts or dyads of non-experts; importantly, two expertly trained individuals working together do not appear to be more accurate than one. This suggests interaction, whereby dyads are allowed to exchange personal information, may inherently give rise to as-of-yet unexplored psychological biases that hurt performance, even when there is equality in the distribution of knowledge.

To explain our findings we theorize the following scenarios:
\begin{enumerate}[label=(\alph*)]
\item {\bf Process losses due to social dynamics:} our findings are in line with literature that has found no pair advantage when the task context consists of general knowledge-based tests involving discrete choice decisions. Theories of social dynamics that could help explain why this is the case, relate to the “process losses” that can plague group settings (e.g., groupthink). One possible explanation for the differential gains in confidence observed across dyad types is with regard to the ability of group settings to reduce individual feelings of uncertainty. Social comparison theory \citep{Festinger1954} posits that individuals are motivated to assess the validity of their opinions by comparing them to those held by others, in the absence of other non-social, “physical” means for doing so. Social comparison processes are thus more likely to operate in groups performing tasks that are more judgmental, as compared with intellective, in nature \citep{Laughlin1982}, as when individuals in a jury scenario adopt a higher threshold for finding a suspect guilty (“beyond a reasonable doubt”) after participating in a group discussion \citep{Magnussen2014,Schuldt2017}.

\item {\bf Overconfidence/equality bias:} overconfident (but inaccurate) people convince less confident (but accurate) people to change their opinions towards the wrong decision; this idea is regularly invoked in the field of collective decision making and is advanced, for instance, by \cite{valeriani2017}. \cite{blanchard2020}, for instance, argue that overconfidence may be responsible for the failure to improve group decisions, as they find that, on average, dyads became more confident than individuals despite no increase in accuracy. The reasons they give to explain this, relate to the metacognitive constructs (i.e., trait-confidence and bias), that is, the presence of one cautious individual (i.e. lower trait-confidence and underconfident) may cause unjustified increases in confidence di-played by dyads composed of overconfident and potentially higher trait-confidence individuals. In a different but related vein, \cite{mahmoodi2015} argue that the reason individuals in a dyad weigh each others' opinions equally, even if one individual is unjustly confident, is because there is an “equality bias” across cultures, meaning when making decisions together we tend to give everyone an equal chance to voice their opinion. In our experiment, this may be especially useful for explaining why mixed dyads performed poorly.

\item {\bf Insufficient time for discussion/social learning theory:} in our experiments time was also a  practical constraint. The time pressure of completing the task in a 45-minute window meant dyads failed to properly discuss and instead the more confident/dominant individual was able to push their opinion through more easily. The more theoretical reason that we could use to explain why time matters, relates to social learning theory (see, e.g., \cite{toyokawa2019}—‘learning that is facilitated by observation of, or interaction with, another individual or its products, is known as ‘social learning’ \citep{kendal2018}. Despite the positive connotation of the term learning, another face of social learning is herding effects where the collective decision gradually moves toward a less optimal solution \citep{vives1996}.

\end{enumerate}

In a  situation where a general manager is deciding whether or not to rely on a qualified expert or a pair of untrained team members for solving various tasks effectively, the results of the present study would suggest that, at least for complex classification tasks, overall performance can be maximized by relying on a pool of expertly trained individuals working alone. Similarly, if the speed at which the decision needs to be made is key, it is also best to rely on individuals working alone. Yet, if accuracy carries more practical importance than productivity, and it is not possible to provide specialized training, then recruiting experts and building mixed-ability teams may be the most effective. 

Providing specialized individual training to all or at least some workers and relying on group work only when accuracy matters may be the most effective strategy, whilst relying on a dyad of trained experts will likely represent a waste of resources, as it does not provide any additional performance gains. Taken together, these results both complement prior findings \citep{Schuldt2017,Bahrami2012a} and provide new insights, highlighting that the extent of training received by an individual, the complexity of the task at hand, and the desired performance outcome are all critical factors that need to be accounted for when weighing up the benefits of collective decision-making. 

Despite the fact that collective decision-making has been studied extensively \citep{Heath1995,Bose2017}, prior studies have relied mostly on artificial or comparatively simple tasks (e.g.\ number line or dot estimation), which do not reflect the nature of human interaction in daily life nor account for the uncertainty and limited task-relevant knowledge that individuals often posses (see Supplementary Material Fig.~S8A, which demonstrates that, in the present case, 86\% of participants had close to none, basic, or average zoology knowledge and zero had any expertise). The Wildcam Gorongosa task studied here, however, is not only an established citizen science challenge engaging thousands of participants but a task that shares significant features with other forms of online collaboration and information processing activities more broadly, given it requires both perceptual ability and general knowledge. Thus, it can be expected that the present results can be generalized to similar environments where collaboration may be substituted or complemented with specialized training to improve outcomes. 

The finding that individuals and dyads with similar training do not perform significantly differently suggests that the pair advantage observed in highly controlled experimental studies employing one-dimensional tasks may be less observable in many multi-dimensional contexts. Yet, given some remaining shortcomings, this study also provides building blocks for future research that can help resolve some of the gaps in our understanding of the relationship between task complexity and the performance benefits experienced by interacting individuals.

A limitation of this study is the fact that only one type of task context was studied. Nevertheless, by focusing on a real-world task and advocating for a ``solution-oriented" approach \citep{Watts2017}, we hope that our approach will inspire further research in computational social science on decision-making using externally valid domain-general task contexts. Moreover, we note that the use of a preexisting citizen science task resulted in a significant proportion of participants reporting an interest in contributing further to citizen science, independent from working alone or in a dyad (see Supplementary Material Fig.~S9 which shows that 68\% of participants reported it as either somewhat or extremely likely that they would contribute to citizen science in the long term), thereby demonstrating the additional benefit to adopting such a task, that is, engaging participants as well as advance our understanding of collective decision-making. 

We believe our study will spur further experimental research using citizen science platforms; we especially encourage researchers to build on our task complexity definition and develop a standardized framework to enable reliable comparisons across citizen science task environments. In the context of collective decision-making research, in particular, when considering individuals with similar expertise who disagree, future studies will advance our understanding of the value of deliberation. Such studies might also vary the time that individuals and groups have to make classifications. Another way of improving on this work is to consider and monitor different strategies that team members take or the difference in features on which they base their decisions. Alignment or misalignment between these could significantly influence team performance.

Finally, we welcome recent calls for the field to take greater inspiration from animal research \citep{OBryan2020} and for researchers to consider how technologies and machines affect human interaction in online collaborative tasks \citep{Rahwan2020} as additional avenues of research to explore and further advance our collective knowledge of collective intelligence.

\section*{Materials and Methods}
This study was reviewed and approved by the Oxford Internet Institute’s Departmental Research Ethics Committee (DREC) on behalf of the Social Sciences and Humanities Inter-Divisional Research Ethics Committee (IDREC) in accordance with the procedures laid down by the University of Oxford for ethical approval of all research involving human participants. Participants ($n=195$) were recruited from the general public via the Nuffield Centre for Experimental Social Science (see Supplementary Material, Sec.~S2). 

\subsection*{Zooniverse Answers}
As all images seen by participants have already been evaluated on the Zooniverse website, Zooniverse estimation data was used to assess performance. These data consist of the aggregated guesses of citizen scientists who used the platform prior to when the experiment was conducted. Whilst volunteer estimates have consistently been found to agree with expert-verified wildlife images \citep{Swanson2016}, it is noted that the ground truth data could still be missing correct attributes that only experts could identify. However, it is by definition impossible for any participant to have performed better than the citizen scientist-provided estimates; hence, we refer to these data as the `ground truth' throughout (see Supplementary Material, Sec.~S3.1 for details on the content of the ground truth).

\subsection*{Performance Analysis}
To comprehensively compare individual versus collective classifications, we assess performance at a specific point and as a function of time. As our primary analysis, we analyze differences in performance change during the final phase of the testing stage. As a secondary analysis, we also examine how any differences evolve over the course of the entire testing stage. In order to do so, we split $T_1$ and $T_2$ into three equally spaced and non-overlapping intervals of length 10 and 15 minutes, respectively. The reason we opted for computing intervals, as opposed to examining performance trial by trial, for example, was because the aim of the study was not to assess the absolute improvements in performance but to compare any change against initial performance during training---before participants were able to acquire any learned expertise. 

Benchmarking changes in performance against average performance during this initial interval in turn ensures we can more precisely estimate whether individual training, interaction, or both, provide performance gains, and how this changes across tasks and over time. We used two different time intervals because each stage was different in length and so we chose the highest factor that would still produce enough data points per stage to allow for at least two comparisons. Next to fulfilling the above criteria, the specific interval sizes were in turn chosen for clarity of presentation, as the results did not significantly change any of the reported findings when changing the size of the interval (see Supplementary Material Fig. S7).

The average change in performance was estimated separately for each interval by computing the difference in performance compared to the performance of the respective individual(s) in the first interval of $T_1$, which can be thought of as the natural baseline prior to training. In comparing individuals to dyads, two benchmarks are used: (a) a ``non-interacting nominal dyad'', defined as the average of the initial performance of both dyad members working individually (the sum is used for volume and efficiency, as these are additive processes), and, in the case of Mixed Dyads, (b) a ``comparably trained individual'', defined as the average initial performance of the dyad member with the same level of training as the individual in the respective solo condition (multiplied by two when considering volume and efficiency). 

Given the sample size, we opted to employ analysis of variance (ANOVA) to statistically qualify the results, instead of applying growth curve modeling, for instance. Similarly, given the length of the experiment, as well as the non-stationary nature of our temporal data, we considered other time-based analysis methods, such as a sliding window approach using a distinct or overlapping time interval, unsuitable as most assumptions would not be met.

\subsection*{Statistical Tests}
All statistical tests were two-tailed and non-parametric alternatives were planned if the data strongly violated normality assumptions. For omnibus tests, the significance of mean differences between groups was analyzed via planned pairwise comparison tests. Given the fact that all pairwise comparisons were planned, and taking into account the sample size, length of the experiment as well as the nature of the data, we follow the recommendation that reporting actual p-values leads to fewer errors of interpretation \citep{rothman1990no, saville1990multiple}. This also means we avoid any arbitrary significance cut-off points \citep{Vidgen2016}. As a result, we do not correct p-values but instead ensure that they are situated within the overall findings of the study. Effect size values were meanwhile interpreted in simple and standardized terms according to \cite{Cohen1988}, when no previously described values are available for comparison, and reported using conventions recommended by \cite{Lakens2013}. Details of the statistical tests are in Supplementary Material, Sec.~S3.3. 

\section*{Acknowledgments}
We wish to thank Shaun A. Noordin and Chris Lintott for their assistance and support with the Wildcam Gorongosa site as well as all the contributors to the Wildcam Gorongosa project. We are grateful to Paola Bouley and the Gorongosa Lion Project, and the Gorongosa National Park for sharing the images and annotations with us. We thank Kaitlyn Gaynor for the useful comments on the manuscript and Bahador Bahrami for insightful discussions.

\section*{Funding}
\textbf{Funding:} This project was partially supported by the Howard Hughes Medical Institute. T.Y. was also supported by the Alan Turing Institute under the EPSRC grant no. EP/N510129/1.  

\section*{Authors Contribution}
V.J.S. analyzed the data and drafted the manuscript. M.T. designed and performed the experiments. T.Y. designed and performed the experiments, designed the analysis, secured the funding and supervised the project. All authors contributed to writing the manuscript and gave final approval for publication. 

\section*{Competing interests}
The authors declare no competing interests.

\section*{Data Availability}
The Gorongosa Lab educational platform used to run the `Wildcam Gorongosa task' is hosted by The Zooniverse \href{https://www.zooniverse.org/projects/zooniverse/wildcam-gorongosa}{(www.zooniverse.com)}. Replication data and code are available at the Open Science Framework: \href{https://osf.io/6rcgx/}{https://osf.io/6rcgx/}.  
\normalsize 

\singlespacing

\bibliographystyle{apa}
\bibliography{references}

\label{LastPage}
\newpage
\thispagestyle{empty}
\clearpage


\pagenumbering{arabic}
\begin{center}
\Large  
\textbf{Supplementary Information For:\\The cost of coordination can exceed the benefit of collaboration in performing complex tasks} 

\normalsize
\noindent \newline \\ Vincent J. Straub\textsuperscript{1}, Milena Tsvetkova\textsuperscript{1,2}, and Taha Yasseri\textsuperscript{1,3,4,5}* 
\end{center} 

\footnotesize 
\noindent  \\ \textsuperscript{1}Oxford Internet Institute, University of Oxford, Oxford OX1~3JS, UK \\
\textsuperscript{2}Department of Methodology, London School of Economics and Political Science, London WC2A~2AE, UK  \\
\textsuperscript{3}School of Sociology, University College Dublin, Dublin D04 V1W8, Ireland \\
\textsuperscript{4}Geary Institute for Public Policy, University College Dublin, Dublin D04 N9Y1, Ireland\\
\textsuperscript{5}Alan Turing Institute for Data Science and AI, 96 Euston Rd, London NW1 2DB, UK
\\ \newline
\noindent *To whom correspondence should be addressed: taha.yasseri@ucd.ie  
\normalsize 

\date{}
\thispagestyle{empty}
\raggedbottom
\cleardoublepage
\pagestyle{fancy}
\lhead{}
\chead{}
\rhead{}
\lfoot{}
\rfoot{}
\renewcommand{\headrulewidth}{0pt}
\renewcommand{\footrulewidth}{0pt}
\renewcommand{\tablename}{\textbf{Table.}}
\renewcommand{\figurename}{\textbf{Figure:}}
\pagenumbering{arabic}
\renewcommand\thefigure{\thesection.\arabic{figure}} \setcounter{figure}{0}





\pagenumbering{arabic}
\setcounter{page}{1}

\renewcommand{\thesection}{S\arabic{section}}
\renewcommand{\thetable}{S\arabic{table}}
\renewcommand{\thefigure}{S\arabic{figure}}
\renewcommand{\tablename}{\textbf{Table.}}
\renewcommand{\figurename}{Figure:}

\section{Access to Code and Data}
This study was reviewed and approved by the Oxford Internet Institute’s Departmental Research Ethics Committee (DREC) on behalf of the Social Sciences and Humanities Inter-Divisional Research Ethics Committee (IDREC) in accordance with the procedures laid down by the University of Oxford for ethical approval of all research involving human participants. Replication data and code are available at the Open Science Framework accessible via the following url: \url{https://osf.io/6rcgx}. 

The Gorongosa Lab educational platform is hosted by The Zooniverse (zooniverse.org), an open web-based platform for large-scale citizen science research projects \citep{Cox2015}. All pictures shown are by Wildcam Gorongosa licensed under CC BY 4.0. The survey responses were recorded using the Qualtrics software \citep{Qualtrics2020}. 
\normalsize 

\section{Details of Experimental Setup}
\subsection{Wildcam Gorongosa Classification Task}
For the experiment, participants were asked to solve citizen science classification tasks using the Wildcam Gorongosa platform, first individually during a Training stage, $T_1$, and then potentially in a dyad with another participant during a Testing stage, $T_2$ (i.e., based on whether the participant is assigned to an individual or dyad condition). Citizen science is a form of crowdsourced e-Research whereby volunteering amateurs or nonprofessional scientists collaborate with experts on scientific research \citep{Watson2018}. These so-called `citizen scientists' primarily carry out tasks for which human-based analysis often still exceeds that of machine intelligence \citep{Trouille2019}. In the context of Zooniverse, these simple intelligence tasks include transcription, annotation, and drawing, among others \citep{Rosser2018}. In the present case, the participant(s) were tasked with recognizing animals in pictures taken by camera traps in Mozambique’s Gorongosa National Park set up to document the recovery of wildlife populations (the trail cameras are designed to automatically take a photo when an animal moves in front of them). The tasks participants had to perform are thus real-world problems which constitute an important part of both wildlife research and conservation---Wildcam Gorongosa has engaged over 40,000 volunteers to date \citep{WildcamGorongosa} and is actively used in classroom settings in the life and environmental sciences \citep{Bolen2018}. Whilst object recognition more broadly is considered a complex task which relies on different components and is consequently a subject of intensive research in vision science, artificial intelligence, and human-machine collaboration \citep{Hinton2012,Russakovsky2015,DiCarlo2012}. 

For each image, participants in the experiment were tasked with (1) detecting the presence of the animal(s), (2) identifying the species type, (3) counting how many animals there are, (4) identifying the behaviours exhibited, specifically, identifying whether the animal(s) is (a) standing, (b) resting, (b) moving, (c) eating, or (d) interacting behaviour (multiple behaviours may be selected), and (5) recognizing whether any young are present. The 52 possible species options include a `Nothing here' option and four `group' categories: human, bird (other), reptiles, and rodents. Whilst all images participants saw were classified by citizen scientists as containing animals, the `Nothing here' button allowed participants to still classify images if they failed to detect any visible animals. The option `human' is meant to indicate any human activity, such as the presence of vehicles. Some images contained more than one species (see Sect. S3.1); however, for the species identification task participants could only identify one instead of multiple species (i.e., single-label classification). Similarly, options for the animal count (binned as  1,  2,  3,  4,  5,  6,  7,  8,  9,  10,  11-50,  and  51+  individuals) and detecting young task (options are Yes/No) are also instances of single-label classification. Nevertheless, classifying behaviour can be considered an instance of a multi-label classification task as multiple behavioural attributes can be selected, consequently, this task was evaluated differently to the others (see Sect. S3.2). Participants were able to filter potential species by morphological characteristics, such as body shape (see Fig. S4). 

The pictures seen by participants have already been classified on the Zooniverse website, each one having been classified by at least 25 citizen scientists. All classifications made by citizen scientists for the same picture are aggregated into the most likely answer in order to provide a `correct' label  for each of the tasks described \citep{WildcamGorongosa}. When aggregated this way, the accuracy of citizen scientist-provided labels is on par with labels provided by expert zoologists \citep{Swanson2016}; hence these labels are used in the experiment as the effective `ground truth' to evaluate the decisions of participants (see also Study Design in the main text). For example images and the interface for the classification task that is undertaken by individuals and dyads, see Figs. S1-S2.

\begin{figure}[!ht]
\centering
\includegraphics[width=\textwidth]{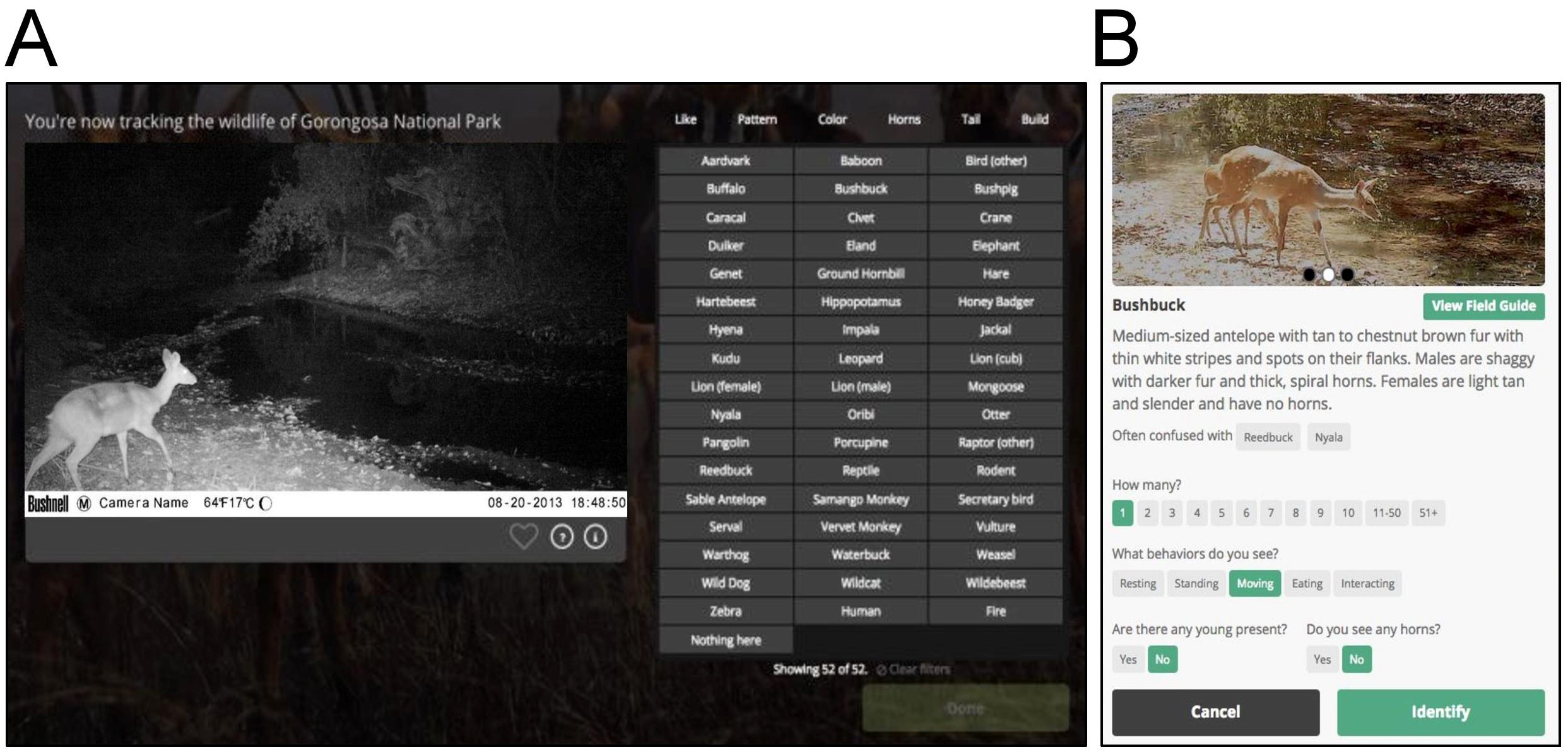} 
\caption{\textbf{The Wildcam Gorongosa interface}. The primary interface (\textbf{A}) asks users to first detect and identify the type of species in the picture, filters (\textbf{B}) then help users identify the number of individuals present, recognize whether any young are present, and identify the behaviours exhibited.}
\end{figure}

\begin{figure}[!ht]
\centering
\includegraphics[width=\textwidth]{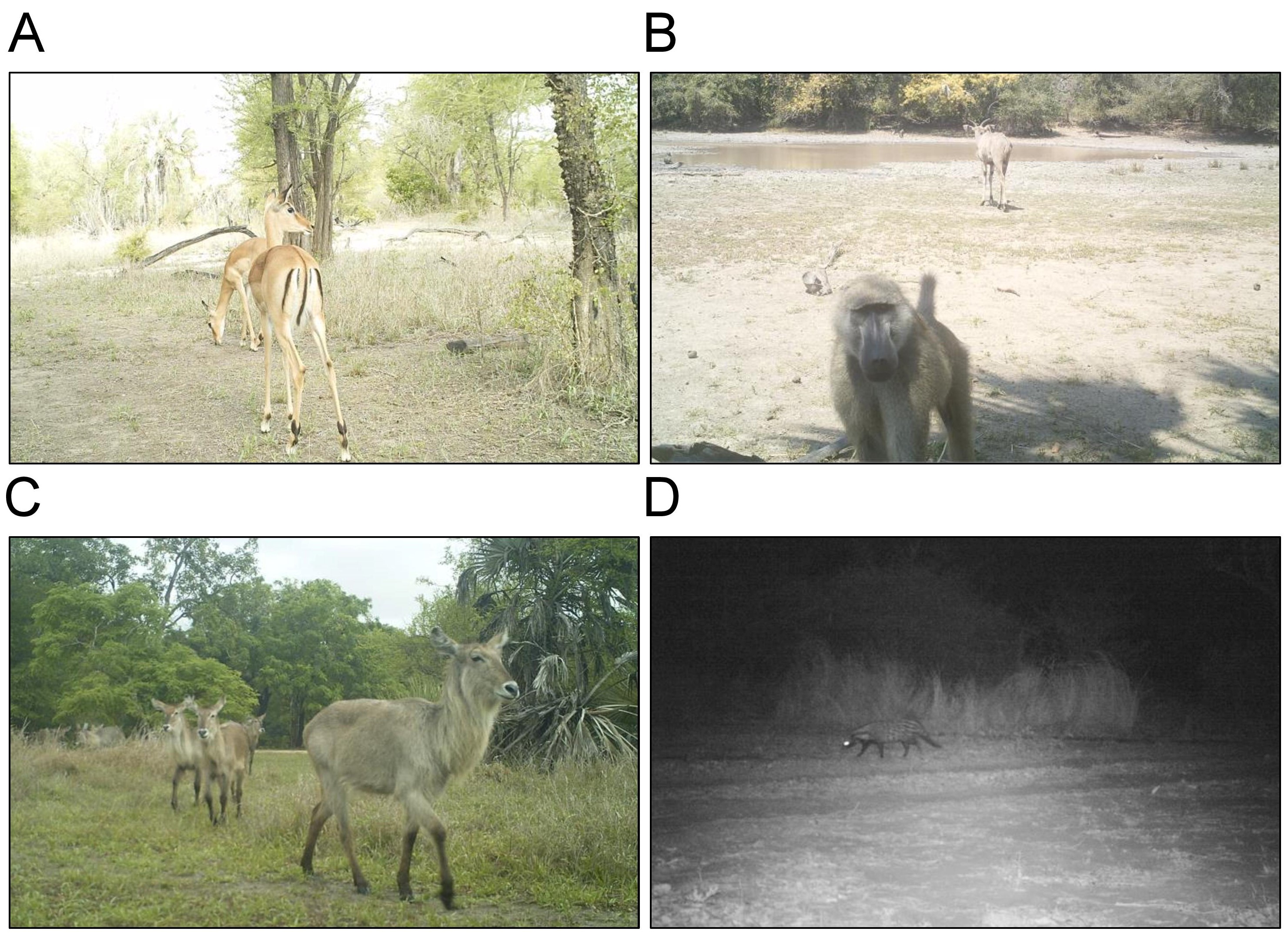}
\caption{\textbf{Example images}. Shown are four images seen by participants in the experiment. Various factors make classifying the images difficult. Detecting and counting the number of animals may be considered tasks of `lower complexity' compared to identifying whether any young are present alongside identifying the behaviours, and identifying the species.}
\end{figure}

\newpage 

\clearpage

\subsection{Screenshots of the Instructions}
\begin{figure}[!hb]
\centering
\includegraphics[width=\textwidth]{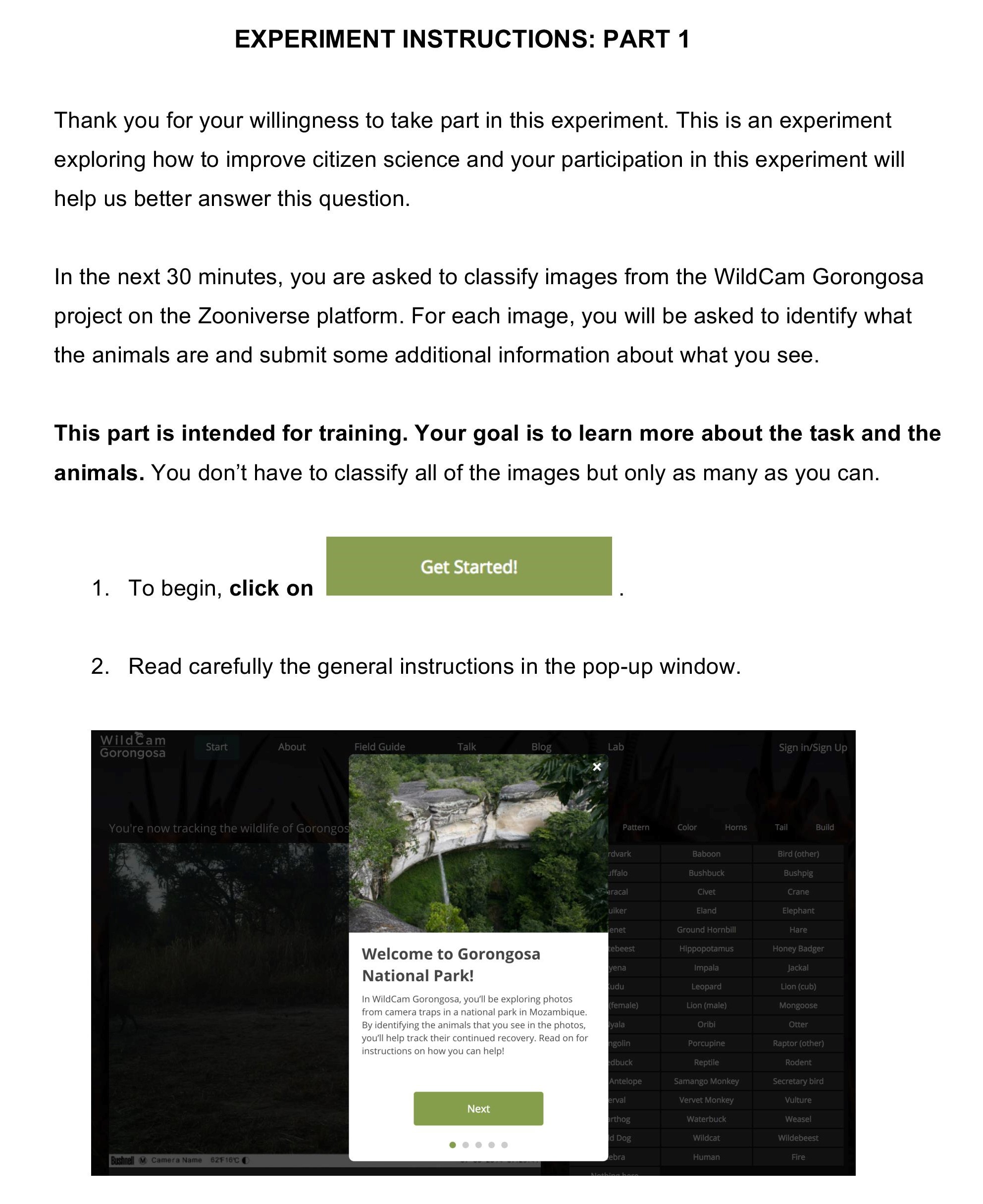}
\caption{\textbf{Experiment instructions.} The Wildcam Gorongosa platform provides users with brief instructions on how to use the interface to classify images.}
\end{figure}

\newpage

\begin{figure}[!ht]
\centering
\includegraphics[width=\textwidth]{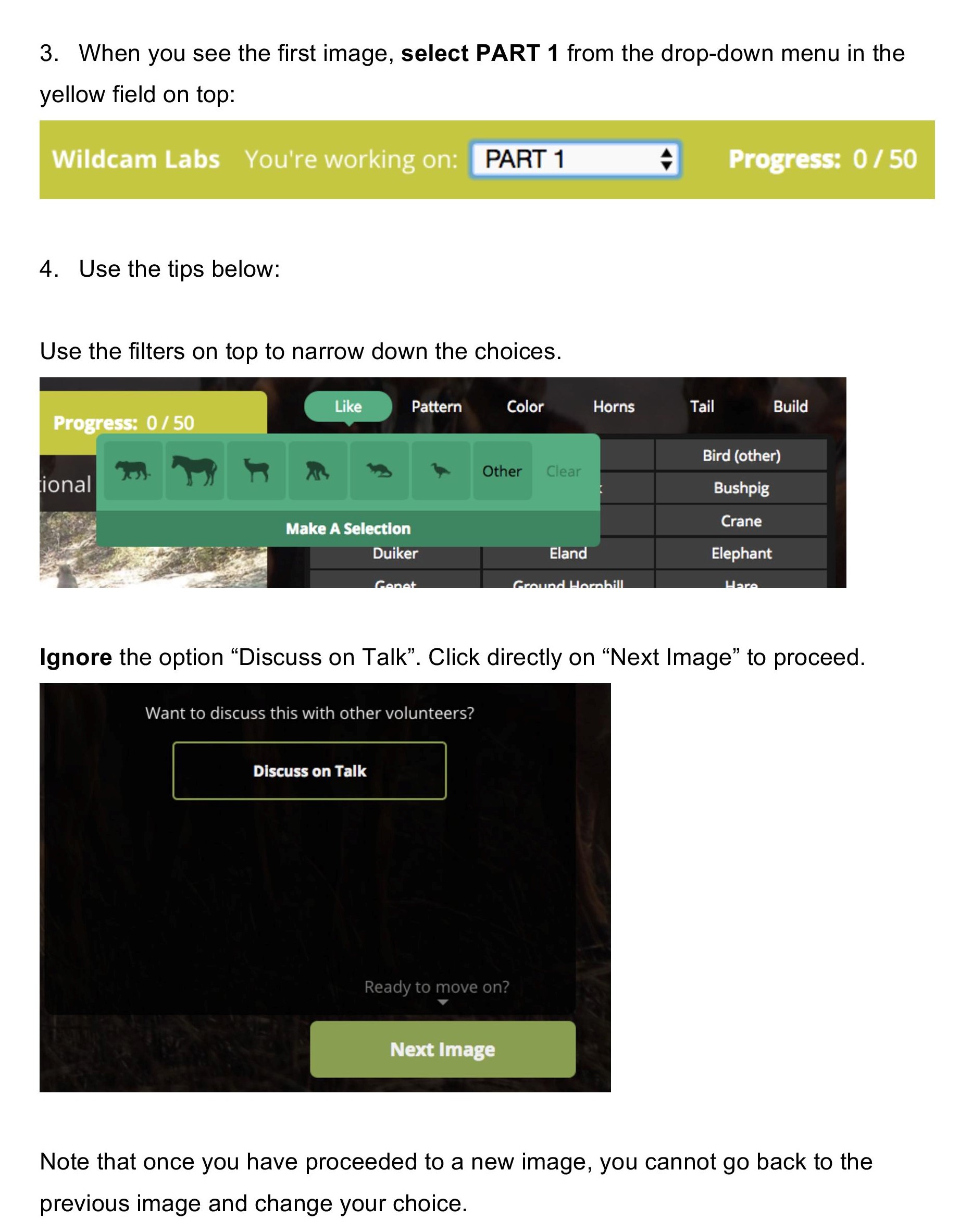}
\caption{\textbf{Experiment instructions.} Users can only classify an image once.}
\end{figure}

\newpage
\newpage
\newpage

\subsection{Experimental Design}
The experiment was conducted at the University of Oxford. Participants were recruited from the general public through the Nuffield Centre for Experimental Social Science (CESS). All participants ($n=195$) provided explicit consent to participate in the experiment and were 18 years of age or older  ($M=28.6$, $SD=12.2$). The experiment was divided into a Training stage, $T_1$, and a Testing stage, $T_2$. In each stage of the experiment, participants first read instructions and could start classifying images only after they had clicked through on each of the instruction slides (see Figs. S3-S4 for screenshots and examples of the instructions given to participants). Participants were in a single room, each facing a large screen; participants working in a dyad were sat in front of a single monitor. 

Two independent variables were manipulated, namely, the type of training received during the training stage (`General' vs. `Targeted') and the grouping condition for the testing stage (`Solo' vs. `Dyad'). Participants in the dyad condition were assigned to form an equal number of `General Dyad' (where both received general training), `Targeted Dyad' (where both received targeted training), and `Mixed Dyad' groups (where only one participant received targeted training). As with allocation to either Targeted or General training conditions for $T_1$, participants were randomly allocated to one of the solo or dyad conditions for $T_2$. Participants had 30 minutes to complete the $T_1$ and 45 minutes to complete $T_2$. During $T_2$, participants working in a dyad did not know each other before the experiment. At the end of testing, participants individually completed an exit survey which asked questions about their demographic background, previous experience and future intentions regarding citizen science, as well as experience of the experiment, discussed below. Participants were paid £18 upon successfully completing both the training and testing stages. 

\subsubsection{Design of the Training Stage}
The training stage was designed to give all participants a better understanding of the interface and tasks, as well as provide selective practice to participants in the Targeted treatment condition ($n=97$) in preparation for the testing stage. The set of images seen by participants in the Targeted training condition (Targeted set) thus consisted of 50 predetermined pictures of animals that are also included in the Testing stage. In contrast, for participants in the General training condition ($n=98$), the set of images (General set) consisted of 50 predetermined pictures of animals that are different from the animals included in the Testing stage. In both training conditions, if participants classified more than 50 images in the time allotted they were able to continue classifying images that were sampled from either the General set or Targeted set, depending on their training treatment condition. Importantly, the exact sequence of images seen by each participant was not artificially controlled. Rather, to strengthen external validity, participants classified images in a way similar to citizen scientist volunteers using the Wildcam Gorongosa site specifically and the Zooniverse platform more broadly.

For the Targeted set, images were restricted to those containing one or more of the following antelope species: bushbuck, duiker, impala, kudu, nyala, oribi, reedbuck, and waterbuck. These animals were chosen as they look similar to each other, share a number of morphological features, and exhibit similar behaviours, thus making them relatively harder to distinguish and ensuring greater variation in classification performance across the various tasks (see Fig. S5). For the General set, images were restricted to those containing: baboon, bird (other), caracal, civet, genet, hyena, jackal, lion (cub), lion (female), lion (male), mongoose, serval, warthog, wild dog, wildcat. At the time of the experiment, there were 14,333 and 20,568 such pictures on Wildcam Gorongosa, respectively. Given that images may contain more than one species and as such be classified more than once by citizen scientists (see Sect. S3.1), a small proportion of the images seen by participants in the targeted condition  ($3.8\%$) also contained baboons, civets, and warthogs, i.e. non-antelope species. Similarly, a small minority of the images ($12.4\%$) seen by participants in the general condition included species belonging to the list of antelope species listed above.  

\begin{figure}[t!]
\centering
\includegraphics[width=0.75\textwidth]{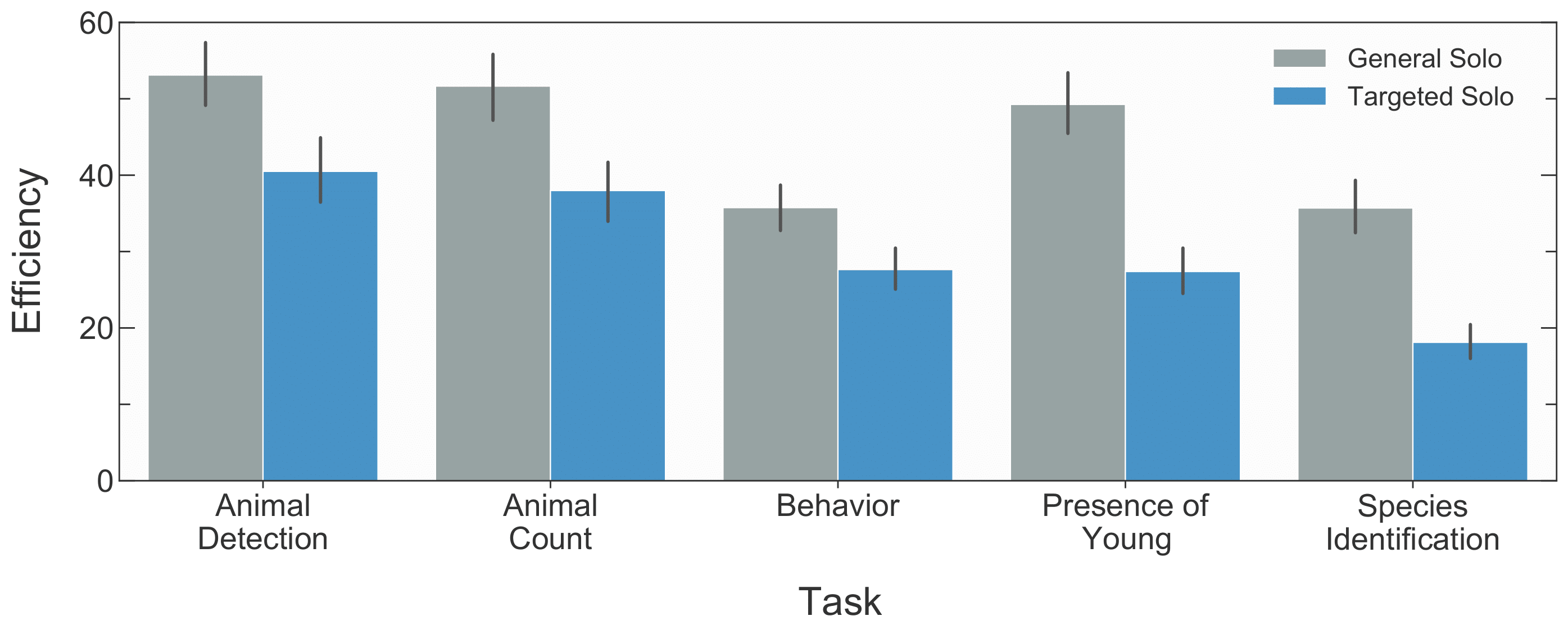} 
\caption{\textbf{Differences in training stage performance.} Shown is performance in terms of efficiency, the number of correct classifications made, during the training stage for each task. Data is combined across individuals for both the Targeted and General training condition. Across all tasks, participants in the Targeted condition on average classified fewer images correctly thereby helping to validate the increased relative difficulty of classifying images in the Targeted set.}
\end{figure}
\subsubsection{Design of the Testing Stage}
After completing the training stage, participants were randomly allocated to one of two additional conditions for the testing stage: Solo vs. Dyad. Specifically, a participant who was assigned to a Solo condition for the testing stage and had been assigned to the General training treatment is referred to as `General Solo' ($n=24$), whilst a participant who was assigned to the Targeted training treatment is referred to as `Targeted Solo' ($n=25$). Similarly, participants in the dyad condition were assigned to form roughly equal numbers of General Dyads ($n=24$), Targeted Dyads ($n=23$), and Mixed Dyads ($n=26$). The set of possible images participants subsequently had to classify was restricted to the same set of pictures as contained in the Targeted set. Moreover, the exact sequence of images seen by participants was again not controlled. Aside from improving external validity, this helped to increase internal validity by ensuring any observed performance differences in the testing stage are robust to slight variations in the set of images seen by participants. In contrast to the training stage, participants were instructed to classify as many images correctly as they could for the testing stage (i.e., there was no predetermined minimum requirement of 50 images). Crucially, participants in a dyad condition had to agree on a joint decision for each image for each of the 5 tasks; one of the participants in the dyad was assigned to input the decisions using a single interface.  

Given all participants had to classify images from the same set of images, whether in a solo or dyad condition, data generated in the testing stage was in turn analyzed in order to address the main research questions of the study.  

After all individuals/dyads finished classifying images for the testing stage, they were asked to individually complete a 5-minute exit survey. Below are the 17 multiple-choice questions for which each participant was asked to provide self-reported answers:
\begin{enumerate}
    \item Age
    \item Gender
    \item Ethnicity
    \begin{itemize}
        \item Bangladeshi, Indian, or Pakistani
        \item Black, African, or Caribbean
        \item Other Asian background
        \item White British or Irish
        \item Other white background       
        \item Other
    \end{itemize}
    \item How well do you know the English language?
    \begin{itemize}
        \item I sometimes find it difficult to understand others or express myself
        \item I am fluent
        \item I am a native speaker
    \end{itemize}
    \item What is/was your subject specialization in school?
        \begin{itemize}
            \item Applied sciences (agricultural sciences, computer science, engineering and technology, medicine and health sciences)
            \item Arts (performing, visual arts)
            \item Humanities (geography, history, languages and literature, philosophy)
            \item Sciences (biology, chemistry, earth and space sciences, mathematics, physics) 
            \item Social sciences (economics, law, political science, psychology, sociology) \item Vocational area (construction, building services, education, accounting, hairdressing, and similar) 
        \end{itemize}
    \item What is the highest educational or vocational qualification that you have or that you will receive if you complete your current degree program?
        \begin{itemize}
            \item Secondary school up to 16 
            \item Secondary school/sixth form college up to 18 with A levels or equivalent
            \item Other college qualification e.g. BTEC, City \& Guilds
            \item University degree or degree level equivalent
            \item Postgraduate degree
        \end{itemize}
    \item Which of the following best describes your current working status?    
        \begin{itemize}
            \item Retired
            \item Student
            \item Working
            \item Unemployed, disabled, homemaker, or other
        \end{itemize}
    \item How would you evaluate your knowledge in the field of zoology before this experiment?
    \begin{itemize}
        \item Close to none (1) --- Expert or equivalent (5) 
            \end{itemize}
    \item What was your experience with citizen science before this experiment?
    \begin{itemize}
        \item I have never heard of citizen science (1) --- I have contributed regularly (4) 
            \end{itemize}            
        \item Please indicate the extent to which you agree with the statement 'I see myself as someone who is talkative':
        \begin{itemize}
            \item Strongly agree (1) --- Strongly disagree (5)  
        \end{itemize}
        \item Please indicate the extent to which you agree with the statement 'I see myself as someone who is full of energy':
        \begin{itemize}
            \item Strongly agree (1) --- Strongly disagree (5)  
        \end{itemize}
        \item Please indicate the extent to which you agree with the statement 'I see myself as someone who generates a lot of enthusiasm':
        \begin{itemize}
            \item Strongly agree (1) --- Strongly disagree (5)  
        \end{itemize}
        \item Please indicate the extent to which you agree with the statement 'I see myself as someone who has an assertive personality':
        \begin{itemize}
            \item Strongly agree (1) --- Strongly disagree (5)  
        \end{itemize}
        \item Please indicate the extent to which you agree with the statement 'I see myself as someone who is outgoing and social':
        \begin{itemize}
            \item Strongly agree (1) --- Strongly disagree (5)  
        \end{itemize}
        \item How useful did you find the first part of the experiment (the training)?
        \begin{itemize}
            \item Not at all useful (1) --- Extremely useful (5)
        \end{itemize}
        \item How difficult did you find the second part of the experiment (the testing)?
        \begin{itemize}
            \item Extremely easy (1) --- Extremely difficult (5)
        \end{itemize}
        \item How much did you enjoy participating in this experiment?
        \begin{itemize}
            \item None at all (1) --- A great deal (5)
        \end{itemize}
        \item How much did you learn by participating in this experiment?
        \begin{itemize}
            \item None at all (1) --- A great deal (5)
        \end{itemize}
        \item How likely are you to volunteer and contribute to a citizen science project in the next few days?
        \begin{itemize}
            \item Extremely unlikely (1) --- Extremely likely (5)
        \end{itemize}
        \item How likely are you to volunteer and contribute to a citizen science project ever again?
        \begin{itemize}
            \item Extremely unlikely (1) --- Extremely likely (5)
        \end{itemize}
\end{enumerate}

\noindent In addition to the above questions, participants who worked in a dyad during the testing stage were asked to provide self-reported answers to the following two questions:
\begin{enumerate}
    \item How would you evaluate the other participant's knowledge of the animals?
    \begin{itemize}
        \item They had worse knowledge than me (1) --- They had better knowledge than me (3) 
    \end{itemize}
\item How would you evaluate the other participant's contribution to the submitted classifications?
\begin{itemize}
    \item They contributed less than me (1) -- They contributed more than me (1)
\end{itemize}
\end{enumerate}

\section{Details of Analysis}

\subsection{Zooniverse Data}
The aggregated citizen scientist-provided labels for each image downloaded from The Zooniverse are used to evaluate the decisions of participants in the experiment. Specifically, these labels are used as the ground truth to evaluate participants on each of the 5 tasks discussed in Section 2 (see also Study Design) for each image. The aggregated classifications provide a single label for each of the tasks. 

For the task of species identification, a small percentage of the unique images seen by participants in the experiment (3.8\%) were labeled as containing multiple species (i.e., `multi-species images'). Consequently, the ground truth dataset contains multiple species entries (i.e., correct answers) for these images. After manually inspecting these images and determining that the distribution of participants who classified multi-species images was evenly split across both training conditions for the training stage and all grouping conditions for the testing stage, a policy of accepting multiple answers was universally adopted in analyzing the data generated by participants. For example, if an image was labelled as containing both nyala(s) and baboon(s) in the ground truth dataset (as is the case for Fig. S2B), both these labels were considered correct when evaluating a participant's decision. 

As mentioned in Sect. S2.1, for the behaviour task, the ground truth dataset also contains labels for five additional behavioural attributes: standing, resting, moving, eating, and interacting. Citizen science identifications for behaviour are not always as accurate as they are for animal count \citep{WildcamGorongosa}. As a result, the data downloaded from The Zooniverse are not aggregated in the same way. For each additional attribute, the data show the ratio of citizen scientists who selected that attribute for the animals in the photo. Because the attributes are not mutually exclusive, multiple behaviours may be present (especially in cases with multiple individual animals)—this was the case for nearly all of the images seen by participants in the experiment (99\%). For example, both ‘eating’ and ‘standing’ may be selected by 25\% of all citizen scientists who labeled the same picture (as is the case for the image shown in Fig. S6A). Given participants in the experiment only had the option to indicate the presence of each attribute using a dichotomous scale, i.e. selecting the behaviour for yes or leaving it blank for no (see Fig. S1), participant's decisions were compared to the proportional scores by treating both as vectors in a multidimensional space and calculating the cosine of the angle between them, as discussed in the following section.  
\begin{figure}[t!]
\centering
\includegraphics[width=\textwidth]{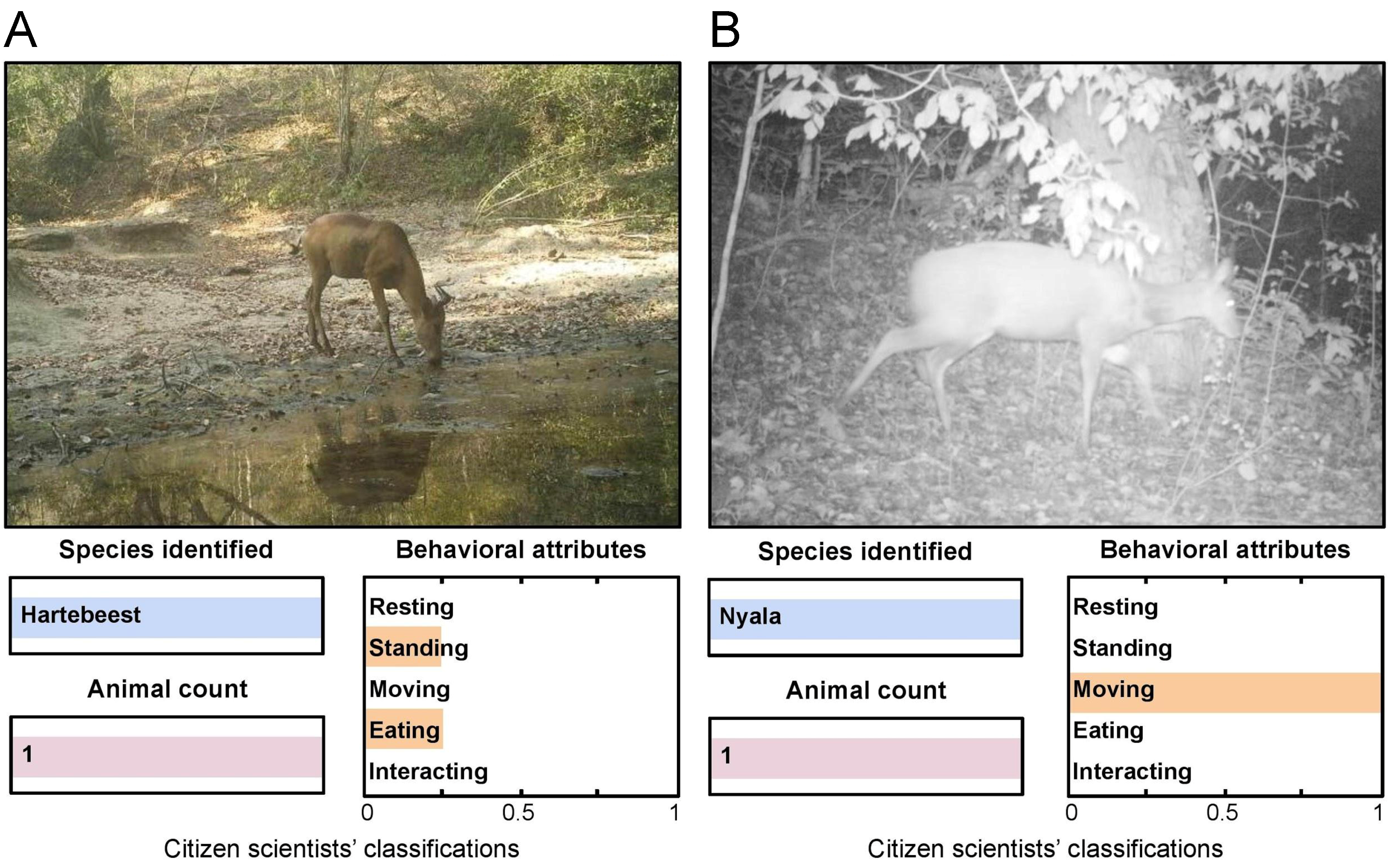} 
\caption{Shown are two images seen by participants in the (\textbf{A}) targeted training (\textbf{B}) and testing stage. Below each image, are the aggregated citizen scientists'-provided classifications for the species and animal count task, as well as the proportions of citizen scientists' classifications for each behavioural attribute. Importantly, although the animal shown in image \textbf{B} is in fact a female Bushbuck, our analysis uses the citizen scientist-provided label of `Nyala' as the effective ground truth value (see S2.1 for details).}
\end{figure}

Further, it is reiterated that the ground truth data could still be missing correct attributes that only experts could identify \citep{Norouzzadeh2018} or contain a few incorrect labels (see Fig. S6B for an example), hence the label `ground truth' is used with care; however, it is by definition impossible for any participant to have performed better than the citizen scientist-provided classifications. Moreover, given that the aggregated citizen scientist-provided classifications provide a single classification for 96.2\% of images seen by participants in the experiment, the label is deemed appropriate and hence used throughout. 

\subsection{Performance Metrics}
As described in the main text, three measures of performance were calculated. Specifically, performance was measured for both the training and testing stage in three ways: 
\begin{itemize}
\item \textit{Volume}: defined as the total number of images classified made by an individual or dyad $i$ (referred to as \textit{pace} when considering images classified per minute)
\item \textit{Accuracy}: defined as the number of correct classifications made by an individual or dyad $i$ as a proportion of volume
\item \textit{Efficiency}: defined as the average number of correct classifications made by an individual or dyad $i$  per unit of time 
\end{itemize}
Except for volume, which is an aggregate measure of performance, accuracy and efficiency were computed separately for each task. In more detail, these metrics were computed as follows. For each image classified by an individual or dyad $i$, a classification was counted as correct in the context of the animal detection task if the species label $x$ did not match the label `Nothing here'. For the species identification and young present task, a classification was counted as correct if the label $x$ selected by $i$ matched the label in the ground truth dataset $y$. For the animal count task, a classification was counted as correct using the sum of the relative difference $d_r$ where the absolute difference between the number $x$ selected by $i$ and the answer $y$ in the ground truth dataset is divided by the maximum absolute value of the two numbers and subtracted from 1: $$d_r = {1 - \frac{|x- y|}{max(x, y)}}$$where $d_r$ takes on a value between
0 and 1 and has the effect of penalizing participants the further away their classification estimate is from the correct value. Although participants have the option to choose a value from the classification interface list [1,  2,  3,  4,  5,  6,  7,  8,  9,  10,  11-50,  and  51+  individuals], as discussed in Section S2.1, the values `11-50' and `51+' are recorded as the numbers 25 and 75, respectively, when downloaded from the Wildcam Gorongosa platform. Finally, for the behaviour task, a participant's decisions $x_{j}$ across each of the 5 behavioural attributes discussed in Sect. S2.1 for an image were collectively compared to the citizen scientist-provided proportional scores $y_j$ by treating both as vectors in a multidimensional space and computing the inner product of the same vectors normalized to both have length 1. This is defined to equal the cosine of the angle between each vector pair which can be used as a measure of similarity between two vectors $sim(\textbf{x, y})$: $$sim(\textbf{x, y}) = \frac{\textbf{x}\cdot\textbf{y}}{| \mathbf{x} || \mathbf{y} | } $$
where $\textbf{x}$ and $\textbf{y}$  are the vectors for a participant's decisions $x_j$ and proportional scores $y_j$ for a single image. In the present case, each vector pair is given a similarity score between $0$ (full dissimilarity) and $1$ (full similarity).

\subsection{Statistical Analysis}
To statistically qualify the results presented graphically in the main text (see Results), we used analysis of variance (ANOVA) to analyze differences in performance during the last third of the testing stage; when the benefits of interaction can best be separated from additional effects, i.e., improvements in coordination. All statistical tests were two-tailed and non-parametric alternatives were planned if the data strongly violated normality assumptions. The significance of mean differences between groups was in turn analyzed via planned tests. Python code for the analysis is provided as a supplementary file in the same repository as the replication data. 

\subsubsection{Analysis of Average Change in Pace}
A one-way ANOVA showed a main effect of grouping condition when comparing General Solo, General Dyad and Mixed Dyad (General benchmark) on pace (Fig. 3A) during the last third of the testing stage (\textit{$F$}(2, 71), = 6.637, $p$=0.002, $\eta_p^2$ = 0.158). Planned comparisons indicated that the average change in pace during the last interval was indeed higher for General Solo ($M$=0.97, $SD$=0.72) compared to General Dyads ($M$=-0.41, $SD$=1.51, $p$=0.001) and Mixed Dyads ($M$=0.01, $SD$=1.62, $p$=0.03), but did not differ significantly between Mixed Dyads and General Dyads ($p$=0.506). 

Further, a one-way ANOVA showed a main effect of grouping condition when comparing Targeted Solo, Targeted Dyad, and Mixed Dyad (Targeted benchmark) (\textit{$F$}(2, 71), = 5.991, $p$ = 0.004, $\eta_p^2$ = 0.144). Planned comparisons similarly indicated that the average change in pace during the last interval was higher for Targeted Solo ($M$=1.65, $SD$=1.04) compared to Targeted Dyads ($M$=0.82, $SD$=0.95, $p$=0.036) and Mixed Dyads ($M$=0.57, $SD$=1.39, $p$=0.003), but did not differ significantly between Mixed Dyads and Targeted Dyads ($p$=0.714).

\subsubsection{Analysis of Average Change in Accuracy for the Species Identification Task}
Results of a one-way ANOVA on the effect of grouping condition when comparing General Solo, General Dyad, and Mixed Dyad (General benchmark) on average accuracy in the species task for the last third of the testing stage show a significant main effect (\textit{$F$}(2, 71), = 4.611, $p$ = 0.013, $\eta_p^2$ = 0.115); planned comparisons further indicate that improvement in accuracy was indeed higher for Mixed Dyads ($M$=0.18, $SD$=0.27) than for General Dyads ($M$=0.00, $SD$=0.16, $p$=0.018) and General Solo ($M$=0.00, $SD$=0.32, $p$=0.036), but did not significantly differ between General Dyads and General Solo ($p$=0.900). 

Similarly, a one-way ANOVA on the effect of grouping condition when comparing Targeted Solo, Targeted Dyad, and Mixed Dyads (Targeted benchmark) on average accuracy in the species task for the last third of the testing stage also shows a significant main effect of grouping condition (\textit{$F$}(2, 71), = 3.652, $p$ = 0.031, $\eta_p^2$ = 0.093); planned comparisons show that improvement in accuracy was greater for Targeted Solo ($M$=0.41, $SD$=0.22) than for Targeted Dyads ($M$=0.24, $SD$=0.28, $p$=0.026), but did not significantly differ between Targeted Solo and Mixed Dyads ($M$=0.29, $SD$=0.24, $p$=0.134), nor between Targeted Dyads and Mixed Dyads ($p$=0.707).

\subsubsection{Analysis of Average Change in Efficiency for the Species Identification Task}

A one-way ANOVA on the effect of grouping condition when comparing General Solo, General Dyad, and Mixed Dyad (General benchmark) on average efficiency in the species task for the last third of the testing stage shows a significant main effect of grouping condition (\textit{$F$}(2, 71), = 3.795, $p$ = 0.027, $\eta_p^2$ = 0.097); planned comparisons show that improvement in efficiency was greater for General Solo ($M$=0.56, $SD$=0.57) than for General Dyads ($M$=-0.23, $SD$=1.05, $p$=0.002), but did not significantly differ between General Solo and Mixed Dyads ($M$=0.27, $SD$=1.24, $p$=0.560) nor between Mixed Dyads and General Dyads ($p$=0.194). 

Finally, a one-way ANOVA showed that the effect of grouping condition when comparing Targeted Solo, Targeted Dyad, and Mixed Dyad (Targeted benchmark) does not have a significant effect on average efficiency for the last third of testing (\textit{$F$}(2, 71), = 1.811, $p$ = 0.171, $\eta_p^2$ = 0.049). This latter statistical result is best interpreted within the context of Fig. 3F which, as discussed in the main text, nevertheless shows that Targeted Solo participants did experience slightly greater efficiency improvements, particularly for the first interval of $T_2$ \citep{Vidgen2016}.

\newpage

\begin{figure}[ht!]
\centering
\includegraphics[width=\textwidth]{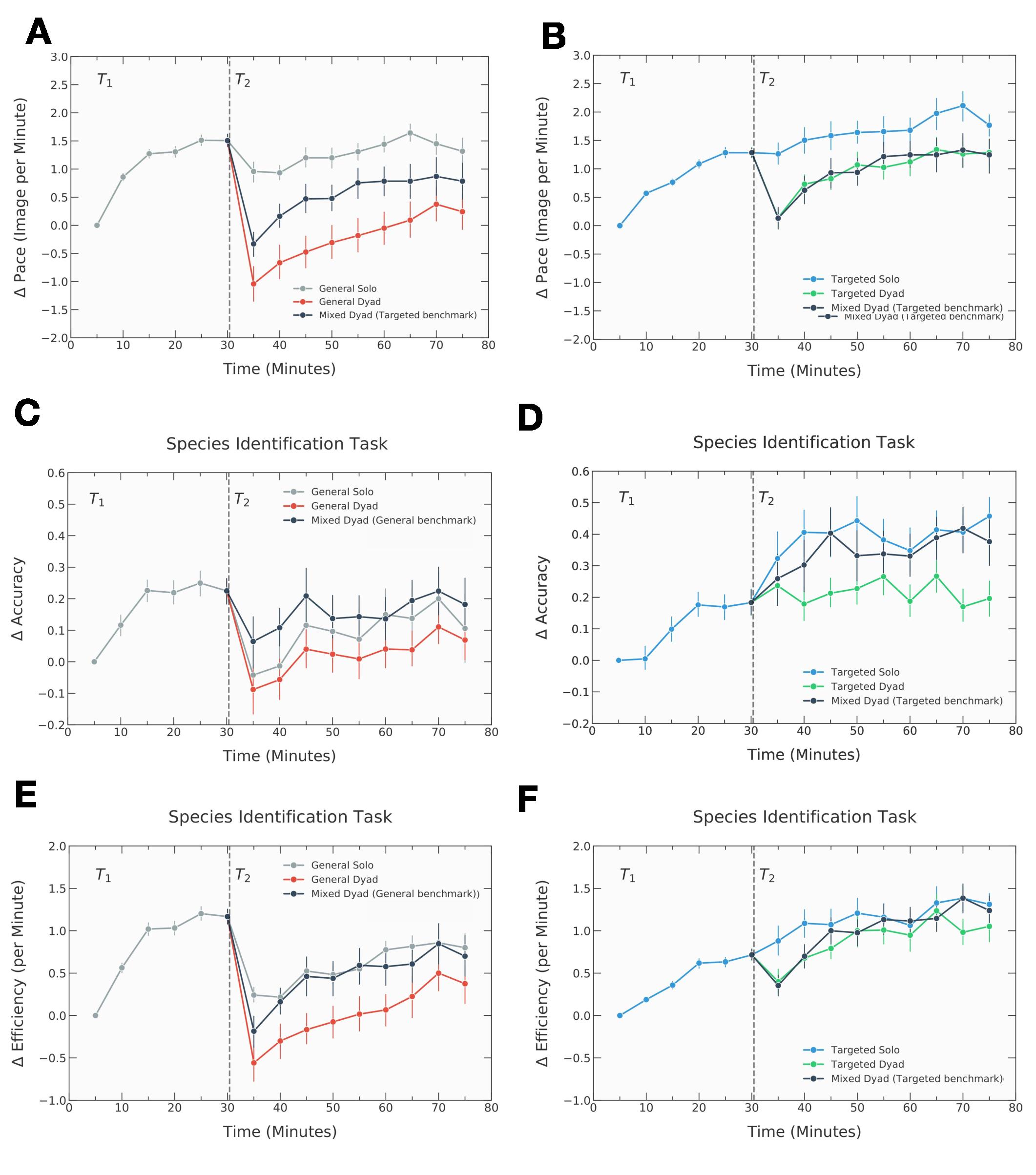} 
\caption{Sliding window analysis using a window with length 5 minutes (equal to 1/6 of total time for $T_1$ and 1/9 of total time for $T_2$) shows that (\textbf{A}) observed differences between General Solo, Mixed Dyads, and General Dyads in terms of the average change in pace do not significantly differ when decreasing the size of the window;  (\textbf{B}) the same can be observed when comparing Targeted Solo, Mixed Dyads, and Targeted Dyads. The dashed vertical line falling in $T_2$ separates both stages, which each have 3 data points. Error bars represent one standard error of the mean.}
\end{figure}

\newpage

\begin{figure}[h!]
\includegraphics[width=\textwidth]{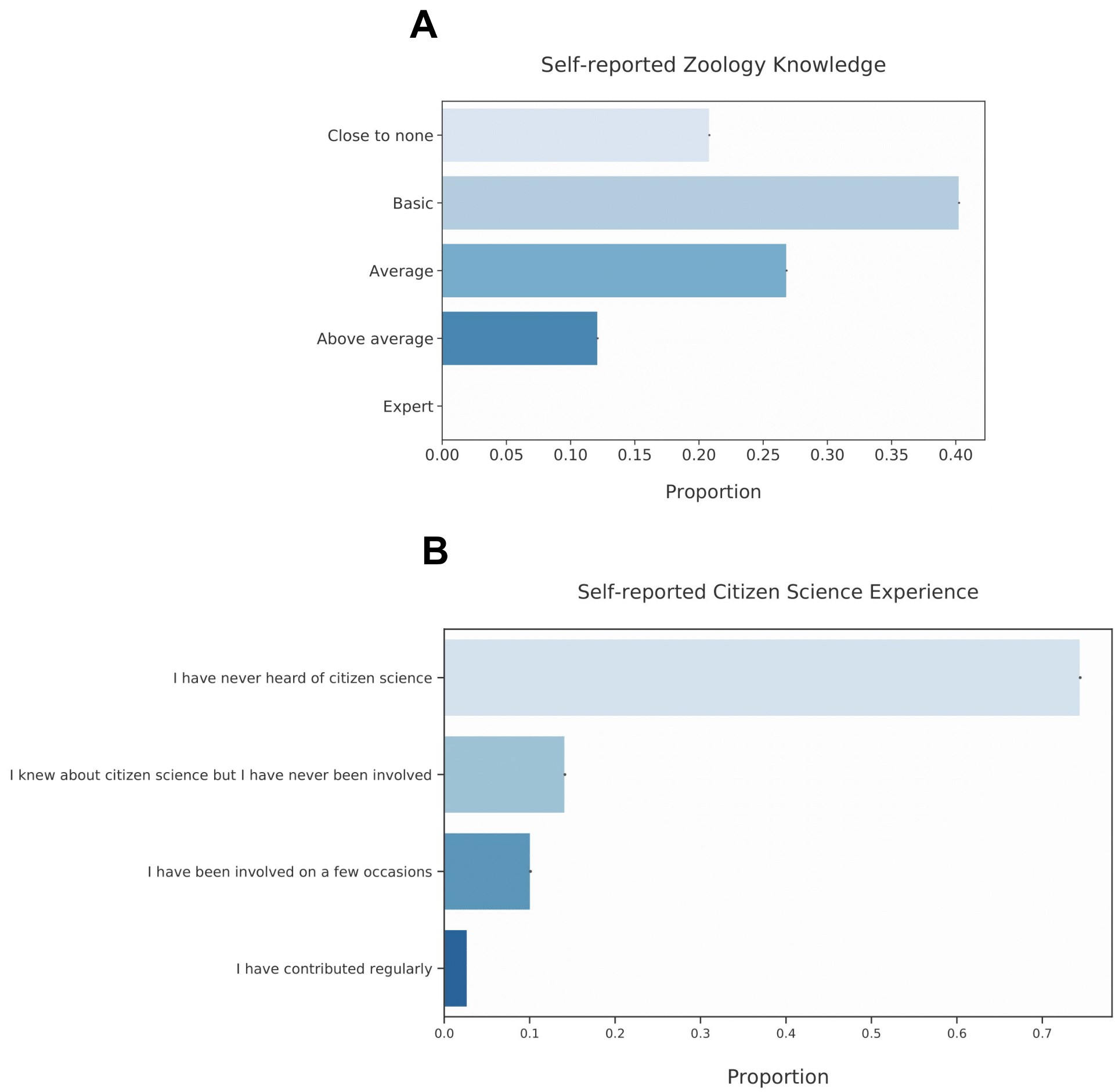}
\caption{Proportion of self-reported answers to Q8 and Q9 of the exit survey completed after $T_2$. No participants reported having expertise in zoology (\textbf{A}); 40\% reported having close to no knowledge, 25\% reported having basic knowledge, 21\% reported having average knowledge, and only 14\% reported having above average knowledge. 70\% of participants reported never having heard of citizen science (\textbf{B}), 16\% had heard of citizen science but not been involved, 12\% had been involved on a few occasions, and only 2\% had contributed regularly (equivalent to one participant).}
\end{figure}

\newpage
\begin{figure}[h!]
\centering
\includegraphics[width=0.8\textwidth]{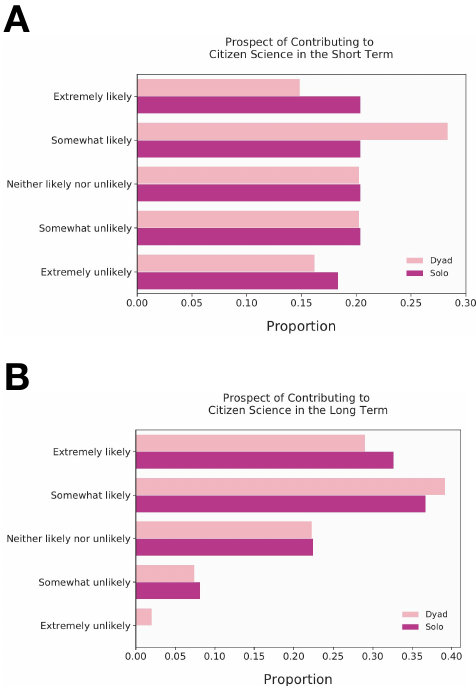} 
\caption{Proportion of self-reported answers to Q19 and Q20 of the exit survey broken down by whether participants worked in one of the Solo or Dyad conditions during $T_2$. Taken together, 43\% of participants who worked in a dyad and 44\% of those who worked individually reported it as either somewhat or extremely likely that they would contribute to citizen science in the short term (\textbf{A}); whilst 68\% and 68\% reported it as either somewhat or extremely likely that they would contribute in the long term, respectively (\textbf{B}).}
\end{figure}

\end{document}